\documentclass[12pt]{article}
\DeclareMathAlphabet{\scr}{U}{rsfs}{m}{n}
\usepackage{supertabular}
\usepackage{amsmath}
\usepackage{amssymb}
\usepackage{graphicx}

\newcommand{\cleqn}{\setcounter{equation}{0}}
\newcommand{\newc}{\newcommand}
\newc{\eps}{\epsilon}
\newc{\Lam}{\Lambda}
\newc{\ra}{\rightarrow}
\newc{\wtilde}{\widetilde}
\newc{\ie}{{\it i.e.}}
\newc{\rpv}{\not\!\! R_p}
\newc{\lsim}{\stackrel{<}{\sim}}
\newc{\beq}{\begin{equation}}
\newc{\eeq}{\end{equation}}
\newc{\beqn}{\begin{eqnarray}}
\newc{\eeqn}{\end{eqnarray}}
\newc{\PLB}{\emph{Phys.Lett.}{\bf{B}}}
\newc{\NPB}{\emph{Nucl.Phys.}{\bf{B}}}
\newc{\mcal}{\mathcal}
\newc{\bsym}{\boldsymbol}
\newc{\nonum}{\nonumber}
\begin{document}
\setlength{\baselineskip}{.6cm}
\title{\textbf{\
Anomalous Flavor $\bsym{U\!(1)_{\!X}}$ for Everything}}
\date{}
\author{~\\Herbi K. Dreiner\footnote{
    E-mail: {\tt dreiner@th.physik.uni-bonn.de}},$^{1}$~
  Hitoshi Murayama\footnote{ On leave of absence from Department of
    Physics, University of California, Berkeley, CA 94720, USA.
    E-mail: {\tt murayama@hitoshi.berkeley.edu}},$^2$~
  Marc Thormeier\footnote{E-mail as of April 2004: {\tt thor@th.physik.uni-bonn.de}} $^{2,3}$
  \\
  \\
  \emph{$^1$Physikalisches Institut der Universit\"at Bonn,}\\
  \emph{Nu\ss allee 12, 53115 Bonn, Germany}\\
  \\
  \emph{$^2$Institute for Advanced Study,}\\
  \emph{Einstein Dr., Princeton, NJ 08540, USA}\\
  \\
  \emph{$^3$Lawrence Berkeley Laboratory,}\\
  \emph{University of California, Berkeley, CA 94720, USA}
}

\maketitle
\begin{abstract}
%
%
  \noindent
\\
  We present an ambitious model of flavor, based on an anomalous
  $U(1)_X$ gauge symmetry with one flavon, only two right-handed
  neutrinos and only two mass scales: $M_{grav}$ and $m_{3/2}$.  In
  particular, there are no new scales introduced for right-handed
  neutrino masses. The $X$-charges of the matter fields are such that
  $R$-parity is conserved exactly, higher-dimensional
  operators are sufficiently suppressed to guarantee a proton lifetime
  in agreement with experiment, and the phenomenology is viable for
  quarks, charged leptons, as well as neutrinos. In our model one of
  the three light neutrinos automatically is massless. The price we have 
  to pay for
  this very successful model are highly fractional $X$-charges which
  can likely be improved with less restrictive phenomenological
  ans\"atze for mass matrices.

\end{abstract}
\newpage
%
%
%
%
%

%
%
\section{\label{intro}Introduction}
\cleqn
   
The fermionic mass spectrum of the standard model ($S\!M$) suggests
that the entries of the quark and charged lepton Yukawa matrices
exhibit hierarchical patterns.  Froggatt and Nielsen (FN) in
Ref.~\cite{fn} presented an idea to explain these texture structures,
 tracing them back to a flavor $U(1)_X$ symmetry beyond the $S\!M$,
which is broken by an $S\!M$-singlet, called the flavon.  For early models see
\emph{e.g.}  Refs.~\cite{lns1,sequel}.  Then string theory gave a theoretical
motivation for the existence and the breaking scale of the $U(1)_X$
(in particular, it turned out that the anomalousness of the $U(1)_X$ is a
blessing), so the FN scenario was easily embedded, see \emph{e.g.}
Ref.~\cite{fma}.
 
One may  naturally  assume that the origin of the aforementioned
hierarchy also leaves its fingerprints on the other (Yukawa) coupling
constants, which opens up many more applications of the FN scenario:  ~\emph{1.)}~It is tempting to
use the idea of FN for investigating why $R$-parity\footnote{$R$-parity was first introduced 
in Ref.~\cite{fafa}.} violating Yukawa
coupling constants have not led yet to the observation of exotic
processes,  see \emph{e.g.} Refs.~\cite{dt,Carone:1996nd}; in these models  $R$-parity violating coupling 
constants are  mostly suppressed down to
  phenomenologically acceptable levels
(rather than them being forbidden exactly),  however  this will
  render the Lightest Supersymmetric Particle not an attractive
  candidate for Cold Dark Matter.   ~\emph{2.)}~Or to provide an explanation
(see \emph{e.g.}  Ref.~\cite{kamu}) why (with or without grand unification)
higher-dimensional genuinely supersymmetric operators like $QQQL$ do
not cause a short-lived proton; after all, these operators are
suppressed by just a single power of the reduced Planck scale, so the
dimensionless coefficient must be adequately tiny.   ~\emph{3.)}~Furthermore, the
idea of FN can easily be combined (see Ref.~\cite{n}) with the
Giudice-Masiero (GM) mechanism (see Refs.~\cite{gm,kn}) to 
naturally explain the  $\mu$-term of the minimal supersymmetric
extension of the $S\!M$ ($M\!S\!S\!M$).    ~\emph{4.)}~Last, but not least, neutrino
data can also be interpreted in the light of the FN idea, see \emph{e.g.}
Ref.~\cite{scheich}. However, when dealing with right-handed neutrinos, the obvious 
question is:  What distinguishes the neutrino superfield from the flavon
superfield?

The aim of this note is to dovetail all of these different aspects of
the FN scenario and at the same time be curmudgeonly about letting
string theory introduce other beyond-$M\!S\!S\!M$ symmetries and/or
particles.  So our intention is to construct a minimalistic
supersymmetric FN model with the following features:
\begin{itemize}
\item There is only one additional symmetry group beyond $SU(3)_C\times
  SU(2)_W \times U(1)_Y$ in the visible sector, namely a
  generation-dependent local $U(1)_X$.  For the cancellation of
  $U(1)_X$ gauge anomalies we invoke the Green-Schwarz mechanism, see
  Ref.~\cite{gs}.  We employ the notation that a left-chiral
  superfield $\Phi$ carries the $U(1)_X$ charge $X_{\Phi}$.
\item There is only one (left-chiral) flavon superfield $A$, which
  acquires a vacuum expectation value (VEV) $\langle A \rangle$ via
  the Dine-Seiberg-Wen-Witten mechanism, see Refs.~\cite{dsw0, dsw,
    ads, ads2}, thus breaking $U(1)_X$.
\item In analogy to the \emph{three} species (\emph{i.e.}  generations) of
  $X$- and $S\!M$-charged matter superfields
  $\{Q^i,L^i,\overline{U^i},\overline{D^i},\overline{E^i}\}$
  ($i=1,2,3$), there are \emph{three} species of superfields whose
  only gauge group is $U(1)_X$: the flavon superfield and two
  right-handed neutrino superfields $\overline{N^I}$ ($I=1,2$).  Thus
  the flavon superfield is so to speak a right-handed neutrino
  superfield without lepton number.\footnote{Having right-handed
neutrinos, we are in no need for neutrino masses which are generated radiatively from $R$-parity violating interactions. We can thus afford to have conserved $R$-parity, which complements the flavon not
carrying lepton number.}
\item The model produces a viable phenomenology: Quark masses and
  mixings and charged lepton masses agree with the data, see
  Refs.~\cite{rrr,n}; the neutrino masses and mixings are in accord
  with the recent measurements, see Ref.~\cite{nirgg} and references
  therein; the lower bounds on proton longevity, see \emph{e.g.}
  Ref.~\cite{prot1,prot2}, and other rare processes are satisfied;
  broken $R$-parity (for a review see \emph{e.g.} 
  Ref.~\cite{rpv2}) is forbidden by virtue of the $X$-charges.
\item There are only two mass scales: the mass of the gravitino
  $m_{3/2}\sim 1~$TeV (we assume gravity mediation of supersymmetry
  breaking, see Refs.~\cite{gramed1,gramed2,gramed3, gramed4}) and the
  reduced Planck scale $M_{grav}\sim2.4\times10^{18}~$GeV where
  gravity becomes strong.\footnote{We simply assume that 
    supersymmetry breaking effects are sufficiently flavor blind, which is possible in  dilaton dominated breaking, or if  the same modular weights are assigned to  
    all three generations (see \emph{e.g.}
    Refs.~\cite{Kaplunovsky:1993rd,Brignole:1993dj}).}$^,$\footnote{For $U(1)_X$ models in supergravity with modular invariance which deal with the soft scalar masses see \emph{e.g.} Refs.~\cite{dudas1,dudas2}.}

\item $\{Q^i,L^i,\overline{U^i},\overline{D^i}, \overline{E^i},
  H^\mathcal{D},H^\mathcal{U}\}$ are the only $S\!M$-charged
  superfields, and we made the (unsuccessful, due to the 
gravitational anomaly) attempt here that those
  fields together with $\{A,\overline{N^I}\}$ possibly are the only
  $U(1)_X$-charged superfields.
\end{itemize}

Our paper is structured as follows: In Section~\ref{scenario} we
review the idea of FN.  In Section~\ref{Consr} we
derive constraints on the $X$-charges such that conserved $R$-parity
is guaranteed.  In Section~\ref{anom} we review the conditions on the
$X$-charges in order to have an anomaly-free theory.  We are then able
to finish the argument started in the Section~\ref{Consr}. In
Section~\ref{pHeNo} we first review the fermionic mass spectrum and
its implications for the $X$-charges. We then combine this with the 
previous results and arrive at Table~\ref{Table2}. Section~\ref{vEv}
discusses how the flavon acquires a VEV (a key ingredient of the
FN scenario).  We also show that   tadpoles do not
endanger our model.  In Section~\ref{uP} we confront the $X$-charges in
Table~\ref{Table2} with further constraints: there are only two mass
scales in the game and attention must be paid to higher-dimensional
operators which destabilize the proton.  Section~\ref{nuPHe} is the
heart of this paper, fixing the $X$-charge assignments by comparison
with neutrino data. A preliminary result is Table~\ref{Table3}, the
main results are given in Tables~\ref{Table5}-\ref{Table8}.
Section~\ref{soc} concludes the paper.  The
Appendices~\ref{seeesooo},\ref{kalk},\ref{mitsusy0} complement
Section~\ref{nuPHe}: reviewing the seesaw mechanism, explaining how
to extract masses from FN textures and including
supersymmetric zeros. Appendix~\ref{bplp} extends Section~\ref{Consr}.

The result of each section is summarized at the beginning.  Hasty
readers can skim through the paper by reading the beginning of each
section together with the tables.

%
%
%
\section{\label{scenario}The  
Framework of Froggatt and Nielsen}
\cleqn

In this section, we review the framework to build models of flavor,
based on a $U(1)_X$ flavor symmetry, which is originally due to Froggatt and
Nielsen (see Ref.~\cite{fn}).  For a review of the FN framework,
possibly combined with the GM mechanism, see \emph{e.g.}  Ref.~\cite{dt}.
Here we shall give only a short sketch.  Models of the same category
as the one constructed in this text are found in
Ref.~\cite{dps,blr,cl,cck,bdls,cch,jvv,mnrv,r,l1,l2,blpr,eir,hm,bs,ky}.\footnote{Examples
of models with \emph{two}  
 flavon superfields of opposite $X$-charges are 
Refs.~\cite{fma,scheich,griechen,ellis+griechen}.}  Note that we do not 
introduce heavy
vector-like fields unlike the original proposal (see Ref.~\cite{fn})
but rather use a simple operator analysis.  We also pay careful
attention to supersymmetric zeros (see below) and how they are filled up by
canonicalizing the K\"ahler potential.

The idea of FN in its simplest form is as follows: One introduces the
above mentioned $U(1)_X$ symmetry and the 
superfield~$A$.\footnote{Since $U(1)_X$ is Abelian, a kinetic mixing term
with $U(1)_Y$ is generically possible. However, as seen below, $U(1)_X$
is broken at a high energy scale. For the low-energy effective theory
the $U(1)_X$ gauge field can be integrated out and thus none of the 
discussions below is affected by the presence of the kinetic mixing.}
 Above the
$U(1)_X$ breaking scale, \emph{e.g.} the coupling constants ${G^{(U)}}_{\!ij}$ of the 
$M\!S\!S\!M$ superpotential Yukawa interaction term
\beq\label{hinz} 
{G^{(U)}}_{\!ij}~~ Q^i~ H^\mathcal{U}~
\overline{U^j}
\eeq 
are promoted to
%
\beq\label{huns}
\Theta\big[X_{Q^i}+X_{H^{\mcal{U}}}+
X_{\overline{U^j}}\big]~\cdot~\Omega\big[
X_{Q^i}+X_{H^{\mathcal{U}}}+
X_{\overline{U^j}}\big]~{g^{(U)}}_{\!ij}~~\bigg(\frac{A}{M}
\bigg)^{{X_{Q^i}+X_{H^\mathcal{U}}+X_{\overline{U^j}}}} ,
\eeq
with \mbox{$X_A=-1$}. 
%
The powers of $A$ in Eq.~(\ref{huns}) compensate the $U(1)_X$ charges
of fields in Eq.~(\ref{hinz}) to form $U(1)_X$ gauge invariants.
$M$ is a high mass scale above which new physics occurs, 
${g^{(...)}}_{\!ij}$ and (see below) ${h^{(...)}}_{ij}$, 
$\widetilde{g^{(...)}}_{ij}$, $\gamma_{IJ}$, $\psi_{ij}$, 
$\widetilde{\gamma}_{IJ}$, $\widetilde{\psi}_{ij}$ are  
dimensionless coupling constants of $\mathcal{O}(1)$, \emph{i.e.}\footnote{Of course it is rather arbitrary to define ``$\mathcal{O}(1)$'' as in  Eq.~(\ref{Range}), stemming from the experimental result that we have two hands,
 each with five fingers. One gets a stricter ~--~however equally
arbitrary~--~ definition with
Eq. (\ref{eeppss}): $
~{\sqrt{\eps~}}\lesssim |...| \lesssim 
\frac{1}{\sqrt{\eps~}}.$}   
\begin{equation}\label{Range}
\frac{1}{\sqrt{10~}}\lesssim |...| \lesssim 
\sqrt{10~}\,;
\end{equation} 
furthermore
\begin{eqnarray}
{\Theta}[x]&\equiv&\left\{
\begin{array}{cl} 
1 & \;\mbox{for }x\geq0  \\
0  & \;\mbox{else (``supersymmetric zero'')} 
\end{array}\right., \\
\label{cdp}
{\Omega}[x]&\equiv&\left\{
\begin{array}{cl} 
1 & \;\mbox{for }x \in\mathsf{Z}\!\!\mathsf{Z}  \\
0  & \;\mbox{else} \end{array}\right..
\end{eqnarray}
${\Theta}[x]$ arises because the superpotential is a holomorphic
function (thus it contains no right-chiral superfields),
${\Omega}[...]$ expresses the fact that the interaction Hamiltonian
density must be a power series of field operators in order to satisfy
the cluster decomposition principle, (\emph{i.e.} distant experiments give
uncorrelated results; Ref.~\cite{wichmann}), see
Ref.~\cite{weinbergtalk} and Chapters~4 and 5 of
Ref.~\cite{weinbergbook}.\footnote{Non-perturbative interactions can
  generate fractional exponents, see Ref.~\cite{br}.  However, such
  effects arise together with a new dynamical scale $\Lambda$, and go
  against our minimalist approach to construct a theory based only on
  two mass scales.  We do not consider this possibility further in
  this paper.}

After $A$ acquires a VEV one has, with
\beq\label{eeppss}
\epsilon\equiv\frac{\langle A\rangle}{M},
\eeq
that effectively
\beq
{{G^{(U)}}_{\!ij}}~~=~~\Theta[X_{Q^i}+X_{H^{\mcal{U}}}+
X_{\overline{U^j}}]~\cdot~
\Omega[X_{Q^i}+X_{H^{\mcal{U}}}+
X_{\overline{U^j}}]~\cdot~{g^{(U)}}_{\!ij}~~
\epsilon^{X_{Q^i}+X_{H^\mathcal{U}}+X_{\overline{U^j}}}.
\label{eff} 
\eeq
Several features of this construction are worth emphasizing:
\begin{enumerate}
\item The other trilinear superpotential terms are obtained the same
  way.
\item\label{hidim} Higher-dimensional (non-renormalizable) operators
  like $Q^iQ^jQ^kL^l$ are obtained analogously, suppressed by powers
  of $M$.
\item Kinetic terms, \emph{i.e.} the bilinear terms of the K\"ahler potential,
  are given by (the complex couplings ${h^{(...)}}_{\!ij}$ form a
  positive-definite Hermitian matrix)
\begin{eqnarray}\label{184}
{H^{(Q)}}_{\!ij}~~\overline{Q^i}~Q^j ~\equiv~\Omega
\big[ X_{Q^i}-X_{Q^j}\big]~\cdot~
{h^{(Q)}}_{\!ij}~~\eps^{|X_{Q^i}-X_{Q^j}|}~~
\overline{Q^i}~Q^j.     
\end{eqnarray}  
\item The bilinear $\mu$-term is determined as [$M^{(\mu)}$ is another
  mass scale, see however the next item and the following discussion],
\beq\label{eff2}
\Theta[ X_{H^\mathcal{D}}+X_{H^\mathcal{U}}  ]~\cdot~
\Omega[X_{H^\mathcal{D}}+X_{H^\mathcal{U}}]~\cdot~
M^{(\mu)}~~{g^{(\mu)}}~~\epsilon^{X_{H^\mathcal{D}}+
X_{H^\mathcal{U}}}~~H^\mathcal{D}~H^\mathcal{U}.
\eeq

\item There are other contributions to Eqs.~(\ref{eff},\ref{eff2})
  produced by the breaking of supersymmetry from $D$-terms (GM
  mechanism) which are particularly important if the operators vanish
  due to supersymmetric zeros (\emph{e.g.},
  $X_{H^\mcal{D}}+X_{H^{\mathcal{U}}}<0$ in the above example).  
\end{enumerate}

Let us elaborate more on this last item because it is 
particularly important for the following.
One supposes a left-chiral, $X$-uncharged  hidden-sector  superfield 
$Z$.  This allows us to write $D$-term operators of the form (with  ${\cal F}$ being a holomorphic
function)
\begin{equation}
  \int d^2\theta\ \ d^2\overline{\theta}\ \ \frac{\overline{Z}}{M} \left[
    \Theta[-X_{\cal F}]\ \Omega[-X_{\cal F}]
    \left(\frac{\overline{A}}{M}\right)^{-X_{\cal F}}
    + \Theta[X_{\cal F}]\ \Omega[X_{\cal F}]
    \left(\frac{A}{M}\right)^{X_{\cal F}} \right]  {\cal F}.
\end{equation}
Decreasing the energy scale,  first $U(1)_X$ is hidden, then 
supersymmetry is broken by the $F$-component of $Z$ acquiring
 a VEV,   which projects out additional  superpotential terms.  Assuming 
 gravity mediation of supersymmetry breaking, such that $\langle F_Z \rangle \sim m_{3/2} M_{grav}$, we get
\begin{equation}
  \int d^2\theta\ \frac{m_{3/2}\ M_{grav}}{M} \ \ \Omega[X_{\cal F}]\
  \Big[ \Theta[-X_{\cal F}]\
    \overline{\epsilon}^{\ -X_{\cal F}}
    + \Theta[X_{\cal F}]\ \epsilon^{X_{\cal F}}
    \Big]\  {\cal F}.
\end{equation}
Even if the $U(1)_X$ charge of the superfield operator ${\cal F}=\Phi_1\Phi_2...\Phi_n/M^{n-2}$ is negative,
the complex conjugate of $\epsilon$ would allow $U(1)_X$ 
invariants.\footnote{Using $U(1)_X$ gauge invariance, we can
  always take $\langle A \rangle$ (and thus $\eps$) real without  loss of generality.}
Trilinear and higher-dimensional terms are highly suppressed due 
to a factor of $m_{3/2}\ M_{grav}/M^2$, while they are
relevant to the $\mu$-parameter, see Refs.~\cite{gm,kn}: 
For example, 
one has
\beqn
{\Omega}
\big[X_{H^\mcal{D}}+X_{H^{\mathcal{U}}}\big]~\cdot~
\frac{m_{3/2}~M_{grav}}{M}~~\widetilde{g^{(\mu)}} 
~~~\eps^{{|X_{H^\mathcal{D}}+X_{H^{\mathcal{U}}}}|}~~
H^\mathcal{D}~H^\mathcal{U},
~~~~~~~~~~~~~~~~~ & &\nonumber\\
 {\Omega}\big[X_{L^i}+X_{H^{\mathcal{U}}}+X_{
\overline{N^j}}\big]~\cdot~\frac{m_{3/2}~M_{grav}}{M^2}
~~\widetilde{g^{(N)}}_{ij}  ~~\eps^{{|X_{L^i}+
X_{H^{\mathcal{U}}}+X_{\overline{N^j}} | }}~~L^i~
H^\mathcal{U}~\overline{N^j}. & &\label{1gdrv}
\eeqn
%
%
The total contribution from $U(1)_X$ breaking to the $\mu$-term is then
\begin{eqnarray}\label{mue}
\mu&=&\Big(M^{(\mu)}~g^{(\mu)}~\eps^{X_{H^\mathcal{D}}+{
X_{H^\mathcal{U}}}}
~~{\Theta}\big[X_{H^\mcal{D}}+{X_{H^\mathcal{U}}}
\big]\nonum\\
&~&~~+~~m_{3/2}~M_{grav}/M~~\widetilde{g^{(\mu)}}~~\eps^{
|X_{H^\mcal{D}}+X_{H^\mathcal{U}}|}\Big)\cdot{\Omega}\big[
X_{H^\mcal{D}}+X_{H^\mathcal{U}}\big].
\end{eqnarray}
%
It is most natural to have $M^{(\mu)}=M=M_{grav}$, thus forcing
one to have $X_{H^\mcal{D}}+X_{H^\mathcal{U}}$ to be $\approx 24\pm1$
(with $\eps\sim0.22$, see below) or
$X_{H^\mcal{D}}+X_{H^\mathcal{U}}=-1, -2<0$.  In the latter
case, $\mu$ is naturally of the same
energy scale as the supersymmetry breaking effects, as desired
phenomenologically.  The supersymmetry breaking contributions to the
trilinear terms can usually be safely neglected, 
see Eq.~(\ref{1gdrv}), while they can be important in the
case of neutrino mass operators, see Ref.~\cite{Arkani-Hamed:2000bq}.

Since the K\"ahler potential from the outset does not have the
canonical form, one must perform a transformation of the relevant
superfields to the canonical basis, see Refs.~\cite{lns1,fkw,blr,dt,jj}
for more details. For example, for the quark doublets one obtains for
the relevant K\"ahler potential term
\beqn\label{firzen}
\overline{Q^i}~{H^{(Q)}}_{\!ij}~Q^j&=&\overline{\Big[
\bsym{
\sqrt{\bsym{D_{\!H^{(Q)}}}~}~
\bsym{U_{\!H^{(Q)}}}~Q}\Big]^i}~\delta_{ij}~\Big[\bsym
{\sqrt{\bsym{D_{\!H^{(Q)}}}~}~\bsym{U_{\!H^{(Q)}}}~Q}
\Big]^j.
\eeqn
$\bsym{D_{\!H^{(Q)}}}$ is a diagonal matrix, its entries 
are the eigenvalues of the Hermitian matrix 
$\bsym{H^{(Q)}}$; the unitary matrix 
$\bsym{U_{\!H^{(Q)}}}$ performs the diagonalization. 
We define the matrix
\beqn\label{3-zehn}
\bsym{C^{(Q)}}\!
\equiv \!\sqrt{\bsym{D_{\!H^{(Q)}}}~}~ 
\bsym{U_{\!H^{(Q)}}}\,.
\eeqn
$\bsym{C^{(Q)}}$ is then absorbed  into $\bsym{Q}$. This 
redefinition affects the superpotential, \emph{e.g.} 
\beqn
 \bsym{G^{(U)}}~\longrightarrow~\frac{1}{\sqrt{
H^{\left(H^\mcal{U}
\right)}~}}~{{\bsym{C^{(Q)}}}^{-1}}^T~\bsym{G^{(U)}}~ 
{\bsym{C^{(\overline{U})}}}^{-1}.
\eeqn
One has
\beq
\Big[\bsym{{{C^{(Q)}}}}^{-1}\Big]_{\!ij}~\sim~
\eps^{|X_{Q^i}-X_{Q^j}|}, ~~\mbox{etc.}
\eeq

There is one unresolved drawback to mention which is generic
to Froggatt-Nielsen models employing the Giudice-Masiero mechanism.
Similar to the expression in Eq.~(\ref{firzen}), there might be present
$D$-term operators of the type $~\overline{Z}Z/{M}^2~\overline{Q^i}
Q^j$.
After gravity mediated supersymmetry breaking and assuming $M=M_{grav}$
one gets a $D$-term of the form $~{m_{3/2}}^2~\overline{Q^i}
Q^j ~\theta\theta\overline{\theta}\overline{\theta}$, inducing non-universal 
and $U(1)_X$-charge dependent contributions to the sparticle soft squared
masses. This potentially causes problems with low-energy FCNCs, and 
is common to all Frogatt-Nielsen models. We expect (or hope?) this 
problem to be solved together with an as yet non-existent proper 
model for supersymmetry breaking.

%
%
%
%
\section{\label{Consr}Conserved $\boldsymbol{R}$-Parity}
\cleqn

In this section, we show that it is possible to obtain conserved 
$R$-parity as an automatic consequence of the $X$-charge assignment.
 Thus $R$-parity is a result of a gauge symmetry, not a discrete
symmetry.

In general, it is desirable (if possible) to choose the $X$-charges
such that superfield operators which give rise to exotic processes are
either forbidden or strongly suppressed. For broken $R$-parity we
shall follow the first path. In this and the next Section we shall for
the purpose of generality treat an arbitrary number of generations
of $\{Q^i,L^i,\overline{U^i},\overline{D^i}, \overline{E^i}\}$ and
$\{\overline{N^I}\}$, \emph{i.e.} not restricting ourselves 
to $i=1,2,3$ and
$I=1,2$.

Consider  a general gauge invariant term of the
$R$-parity violating $M\!S\!S\!M$ with right-handed 
neutrinos   
($\not\!\!R_p$-$M\!S\!S\!M\!+\!\overline{N}$), containing 
$n_{Q^1}$ times the superfield $Q^1$, etc. The 
$n_{...}$  are  non-negative integers if one deals 
with the superpotential, however they may  be negative
in case of the K\"ahler potential due to charge conjugation,
\emph{e.g.} the term $\overline{Q^2}Q^1$ has
$n_{Q^2}=-1$, $n_{Q^1}=1$. The $X$-charge 
of this 
superfield operator is
\begin{eqnarray}\label{xc}
X_{total}&=&\sum_I~ (n_{\overline{N^I}}~ 
X_{\overline{N^I}})~+~
\sum_i~ (n_{L^i}~X_{L^i}~+~n_{\overline{E^i}}~ 
X_{\overline{E^i}}) \nonumber \\
&+& n_{H^\mathcal{D}}~
X_{H^\mathcal{D}}+n_{H^\mathcal{U}}~
X_{H^\mathcal{U}}~+~\sum_i~ 
(n_{Q^i}~X_{Q^i}~+~n_{\overline{D^i}}~ 
X_{\overline{D^i}}~+~n_{\overline{U^i}}~ 
X_{\overline{U^i}}).~~~~~~~~
\end{eqnarray}
The $n_{...}$  are not independent of each other due to  
$SU(3)_C\times SU(2)_W\times U(1)_Y$ gauge invariance.
 They  are subject to the conditions 
(with $n_Q\equiv\sum_{i}
n_{Q^i}$, etc.) 
\begin{eqnarray}
n_Q-n_{\overline{D}}-n_{\overline{U}}&=&3\mathcal{C},
\nonumber\\
n_{H^\mathcal{D}}+n_{H^\mathcal{U}}+n_Q+n_L&=&2\mathcal{W},
\nonumber\\
Y_{H^\mathcal{D}}~n_{H^\mathcal{D}}+~Y_{H^\mathcal{U}}~
n_{H^\mathcal{U}}~+~Y_Q~n_Q~+~Y_{\overline{D}}~
n_{\overline{D}}~+~~~& &\nonumber\\
Y_{\overline{U}}~n_{\overline{U}}~+~Y_L~n_L~+~
Y_{\overline{E}}~n_{\overline{E}}~+~Y_{\overline{N}}~
n_{\overline{N}}&=&0.
\end{eqnarray}
$\mathcal{C}$ is an  integer, $\mathcal{W}$ is a 
non-negative (if one deals with the superpotential) 
integer, the $Y_{...}$ denote 
hypercharges.\footnote{$Y_Q=-1/3~Y_L$, 
$Y_{\overline{U}}=4/3~ Y_L$, $Y_{\overline{D}}= -2/3~  
Y_L$, $Y_{\overline{E}}=-2~Y_L$, $Y_{H^\mathcal{D}}= Y_L$, 
$Y_{H^\mathcal{U}}= -~Y_L$,  $Y_{\overline{N}}=0$.} 
Solving these three equations 
in terms of the numbers of quark superfields
gives
\begin{eqnarray}
n_Q&=&2\mathcal{W}~-~n_{H^\mathcal{D}}-n_{H^\mathcal{U}}-
n_L,\nonumber\\
n_{\overline{U}}&=& \mathcal{W}~-~\mathcal{C}~+~
n_{\overline{E}}~-~n_{H^\mathcal{D}}~-~n_L,\nonumber\\
n_{\overline{D}}&=&\mathcal{W}~-~2\mathcal{C}~-~
n_{\overline{E}}~-~n_{H^\mathcal{U}}\label{gauge}.
\end{eqnarray}

We now state our central assumptions (using $X_A=-1$), which form the basis for the following analysis and which lead to particularly attractive conclusions:
\begin{enumerate}
\item \emph{\underline{All}  superfield operators which 
conserve the $\mathsf{Z}\!\!\mathsf{Z}_2$-symmetry $R_p$ 
each
have an overall integer $X$-charge. For scalar
left-chiral superfields one may use\footnote{Strictly speaking, $R_p$ the way defined here is  {\it matter-parity}\/, because it is independent of the spin of the field. Therefore, $R$-parity as we use it can be a subgroup of a non-$R$-type symmetry such as a flavor $U(1)_X$ gauge symmetry. Matter-parity, just like $R$-parity, 
is  free of anomalies, since it differs from $R$-parity only by a 
spatial rotation  of $2\pi$.}
\beq\label{rparity}
B_p\equiv(-1)^{n_Q-n_{\overline{U}}-n_{\overline{D}}},~~
L_p\equiv(-1)^{n_L-n_{\overline{N}}-n_{\overline{E}}},~~
R_p\equiv B_p\times L_p.
\eeq 
To be more precise, \underline{all} superfield operators
for which $n_Q-n_{\overline{U}}-n_{\overline{D}}+n_L-
n_{\overline{N}}-n_{\overline{E}}$
is even each
have an overall integer $X$-charge.} 
\item  \emph{\underline{All} superfield operators which do 
not conserve $R_p$ 
each have an overall fractional $X$-charge.
To be more precise, \underline{all} superfield operators
for which $n_Q-n_{\overline{U}}-n_{\overline{D}}+n_L-
n_{\overline{N}} -n_{\overline{E}}$ 
is odd each
have an overall fractional $X$-charge.  It follows 
that all  $\not\!\!R_p$ 
 superfield operators are forbidden, $R_p$ is 
thus conserved exactly.}  
\end{enumerate}
One can draw  several conclusions: 
\begin{itemize}
\item The only texture zeros in the sense of 
Ref.~\cite{rrr} that one may have in \emph{e.g.} 
${\bsym{G^{(U)}}}$
are due to $X_{Q^i}+X_{H^\mathcal{U}}+X_{\overline{U^j}}$
 being
negative.
\item Since $Q^1$ has the same $S\!M$ quantum numbers as
 $Q^2$ etc.,  an $R_p$-conserving  superfield operator 
$~Q^1\Phi_1\Phi_2...\Phi_n~$ guarantees that  
 $~Q^2\Phi_1\Phi_2...\Phi_n~$ is $R_p$-conserving, as well,
 etc. 
{}From Point~1. we thus
find that it is necessary that $X_{Q^{2}}-X_{Q^{1}}$
is integer, etc. Thus
\beqn\label{integerdelta}
X_{Q^{2}}&{=}&X_{Q^{1}}+
\underbrace{\Delta^Q_{21}}_{\mbox{integer}},\mbox{~etc.}
\eeqn 
\item For any  $SU(3)_C\times SU(2)_W\times 
U(1)_Y$ invariant superfield operator 
$~\Phi_1\Phi_2...\Phi_n~$ which violates $R_p$, 
one has that 
 $~\Phi_1\Phi_2...\Phi_n\Phi_1\Phi_2...\Phi_n~$
 conserves $R_p$.
 From Point~1. we find that the $X$-charge of 
the latter operator, namely  $2\times\big(X_{\Phi^1}+
X_{\Phi^2}+...+X_{\Phi^n}  \big)$, is integer. Point~2.  
demands that  $~X_{\Phi^1}+X_{\Phi^2}+...+X_{\Phi^n}~$ 
is fractional. \emph{It follows that all $\not\!\!R_p$ 
 superfield operators  
have an  overall half-odd-integer 
$X$-charge.}\footnote{This reasoning is not affected
by \emph{e.g.} on the one hand the superfield operators 
$L^1L^1\overline{E^i}$   being equal to zero
due to $SU(2)_W$ gauge invariance, but on the 
other hand  
$L^1H^\mathcal{D}\overline{E^i}$  not vanishing.}
\item It follows immediately from the previous point that  
    $X_{\overline{N^I}}$ is
{half-odd-integer}; furthermore  let $~L^i\Phi_1
\Phi_2...\Phi_n~$ be $SU(3)_C\times SU(2)_W\times 
U(1)_Y$ invariant and conserve $R_p$, it 
follows that  $~H^\mathcal{D}\Phi_1\Phi_2...\Phi_n~$ 
does not conserve  $R_p$, hence $X_{L^i}-
X_{H^\mathcal{D}}$ is {half-odd-integer}. So in summary
\beq\label{l0.5}
X_{\overline{N^1}},~~X_{L^1}-X_{H^\mathcal{D}}~~=~
\mbox{half-odd-integer}.
\eeq
\item
The $M\!S\!S\!M\!+\!\overline{N}$ is (by definition) 
$R_p$-conserving, we thus have   
\beq\label{trilin}
\left. \begin{array}{l}
X_{Q^1}+X_{H^\mathcal{D}}+X_{\overline{D^1}}\\ 
X_{Q^1}+X_{H^\mathcal{U}}+X_{\overline{U^1}}\\  
X_{L^1}+X_{H^\mathcal{D}}+X_{\overline{E^1}}\\   
X_{L^1}+X_{H^\mathcal{U}}+X_{\overline{N^1}}\end{array}
\right\}~=~\mbox{integer},  
\eeq
and analogously for the other matter superfields, due to
Eq.~(\ref{integerdelta}).
\item From Eq.~(\ref{rparity}) one sees that
\beqn\label{rrpp}
n_L~-~n_{\overline{N}}~-~n_{\overline{E}}~+~n_Q~-
~n_{\overline{U}}~-~n_{\overline{D}}&=&2\mathcal{R}~+
~\varrho.
\eeqn
$\mathcal{R}$ is an integer, $\varrho$ is $0$ or $1$ if $R_p$ is
conserved or broken.
\end{itemize}
We now plug Eqs.~(\ref{gauge},\ref{integerdelta},\ref{l0.5},
$\!\!$\ref{trilin},\ref{rrpp}) into Eq.~(\ref{xc}) and obtain
%
%
%
%
%
\begin{eqnarray}
X_{total}-\mbox{integer}=(3 X_{Q^1} + X_{L^1})
\mathcal{C}-
\mbox{$\frac{\varrho}{2}$}. 
\end{eqnarray}
The l.h.s. of the above equation  has to be half-odd-integer 
or integer
(depending on whether $R$-parity is broken or
conserved) regardless of the integer $\mathcal{C}$, 
so that one finds
\begin{eqnarray}\label{CASE}
3 X_{Q^1} + X_{L^1}~{=}~
\mbox{integer}
\eeqn
is the necessary and sufficient  condition 
[apart from Eqs.~(\ref{integerdelta},\ref{l0.5},\ref{trilin})] 
on the $X$-charges  for  conserved $R_p$. 
We shall see in the next Section that it
does not contradict  conditions of anomaly cancellation 
via the Green-Schwarz mechanism. In Appendix~\ref{bplp} we shall
comment on the analogous calculations for  $B_p,L_p$. Note also
that one obtains  the condition above   if one has no
right-handed neutrinos at all:  Instead of the last condition
in Eq.~(\ref{trilin}) one has to work with $X_{H^\mathcal{U}}
+X_{H^\mathcal{D}}=~$integer.

Apart from $R_p$, 
the only discrete anomaly-free 
gauge symmetry is $B_3$, see Ref.~\cite{ibanezross}.
($R_p$  has a mixed gravitational
anomaly if there are no right-handed neutrinos;
we require them, but only for  two generations, not
three. This can be fixed with $X$-charged hidden sector
matter, which we  have to assume to exist anyway, see the
end of  Subsection~8.1.)
To have $B_3$ instead of  $R_p$  is not very 
attractive in our case, as it allows (just like $B_p$)
 a 
tree level tadpole term, namely the superpotential term 
which is linear in $\overline{N^I}$. In this case  the 
$\overline{N^I}$ acquire  VEVs, thus spoiling the idea 
that  the flavon field alone breaks $U(1)_X$. But is it
  possible to have  $B_3$ 
\emph{together} with  $R_p$
by virtue of the $X$-charges? $B_3$
 transformations act on superfields as
\beqn
\{Q^i, \overline{N^I} \}&
\longrightarrow&\phantom{e^{4\pi i/3}}~
\{Q^i, \overline{N^I}  \},
\nonumber\\
\{\overline{D^i}, H^\mathcal{U} \}
&\longrightarrow&e^{2\pi i/3}~\{\overline{D^i},
H^\mathcal{U} \},  \nonumber\\
\{\overline{U^i}, L^i,\overline{E^i}, H^\mathcal{D} \}
&\longrightarrow&
e^{4\pi i/3}~\{\overline{U^i}, L^i,\overline{E^i}, 
H^\mathcal{D} \},\label{be-drei}
 \eeqn
to be compared with 
the result of $R_p$ transformations:
\beqn
\{H^\mathcal{D}, H^\mathcal{U}\} &\longrightarrow&
\phantom{e^{i\pi}~}
\{H^\mathcal{D}, H^\mathcal{U}\},\nonumber\\
\{Q^i,\overline{U^i},\overline{D^i},L^i,\overline{N^I},
\overline{E^i}\} &\longrightarrow&
e^{i\pi}~\{Q^i,\overline{U^i},\overline{D^i},L^i
\overline{N^I},\overline{E^i}\}.
\eeqn 
With assumptions analogous to Point~1. and Point~2. 
one finds that all $B_3$ 
conserving operators have integer $X$-charges, while 
all $\not\!\!B_3$
 operators have  $X$-charges that are  integer$\pm
\frac{1}{3}$: If an $S\!M$-invariant 
superfield operator $~\Phi_1\Phi_2...\Phi_n~$ 
violates  $B_3$, then
 $~(\Phi_1\Phi_2...
\Phi_n)^3~$
does not. This is incompatible with 
$\not\!\!R_p$ operators, which have
 half-odd-integer $X$-charges.

%
%
%
\section{\label{anom}Anomalies}
\cleqn

In this section, we work out requirements from the anomaly
cancellation via the Green-Schwarz mechanism on $U(1)_X$ charge
assignments and complement the calculation of the previous section.  
 For more details see Ref.~\cite{dt}.

The cancellation/absence of the mixed chiral anomalies of
 $U(1)_X$ with the gauge group of the $S\!M$, itself and 
gravity demands, see \emph{e.g.} Ref.~\cite{maekawa}, 
\begin{equation}\label{cc}
\frac{\mathcal{A}_{CCX}}{k_C}=\frac{\mathcal{A}_{WWX}}{k_W}
=\frac{\mathcal{A}_{YYX}}{k_Y}=\frac{\mathcal{A}_{XXX}}
{3~k_X}=\frac{\mathcal{A}_{GGX}}{24}
\end{equation}
(relying on  the Green-Schwarz mechanism)  and
\begin{equation} 
\mathcal{A}_{YXX}=0.
\end{equation}
The $\mathcal{A}_{...}$ are the coefficients of the
$SU(3)_{C}\mbox{-}SU(3)_{C}\mbox{-}U(1)_{X}$, 
$SU(2)_{W}\mbox{-}SU(2)_{W}\mbox{-} U(1)_{X}$, 
$U(1)_{Y}\mbox{-}U(1)_{Y}\mbox{-}U(1)_{X}$, 
$U(1)_{X}\mbox{-}U(1)_{X}\mbox{-}U(1)_{X}$, 
grav.-grav.-$U(1)_X$, $U(1)_{Y}\mbox{-}U(1)_{X}\mbox{-}
U(1)_{X}$ anomalies, respectively. 
{The factor of $3$ in the third  
denominator in Eq.~(\ref{cc}) is of a combinatorial nature:
 One
 deals with a \emph{pure} rather than \emph{mixed} 
anomaly.} The 
affine/Ka{\v{c}}-Moody levels $k_{...}$ of non-Abelian 
gauge groups have to be positive integers. In terms of the 
$X$-charges one has, see Ref.~\cite{dt},
\begin{eqnarray}\label{ccx}
\mcal{A}_{CCX}&=&\mbox{$\frac{1}{2}$}\Big[\sum_{i}
\Big(2~X_{Q^i}+X_{\overline{U^i}}+X_{\overline{D^i}}\Big)
\Big], \\
\label{wwx}
\mcal{A}_{WWX}&=&\mbox{$\frac{1}{2}$}\Big[X_{H^\mcal{U}}+
X_{H^\mcal{D}}+\sum_{i}\Big(3~X_{Q^i}+X_{L^i}\Big)
\Big], \\
\label{yyx}
\mcal{A}_{YYX}&=&\mbox{$\frac{1}{2}$}\Big[X_{H^\mcal{U}}+
    X_{H^\mcal{D}}
+\mbox{$\frac{1}{3}$}\sum_{i}\Big(X_{Q^i}+
  8X_{\overline{U^i}}+2X_{\overline{D^i}}+3X_{L^i}+6
  X_{\overline{E^i}}\Big)\Big]\cdot4~{Y_L}^2;~~~~\nonumber\\
 \end{eqnarray}
\begin{eqnarray}
\label{xxx}      
\mcal{A}_{XXX}&=&2{X_{H^\mathcal{U}}}^3+2
{X_{H^\mathcal{D}}}^3+\sum_{i}\Big(6{{X_{Q^i}}}^3+
3{X_{\overline{U^i}}}^3+3{X_{\overline{D^i}}}^3+
2{X_{L^i}}^3+{X_{\overline{E^i}}}^3\Big)\nonumber\\
&~&\qquad +{X_A}^3+
\sum_{I}{X_{\overline{N^I}}}^3+
\mcal{A}_{XXX}^{hidden~sector},~~~\\
\label{ggx}      
\mcal{A}_{GGX}&=&2X_{H^\mathcal{U}}+2X_{H^\mathcal{D}}+
\sum_{i}\Big(6{X_{Q^i}}+3{X_{\overline{U^i}}}+
3{X_{\overline{D^i}}}+2{X_{L^i}}+{X_{\overline{E^i}}}\Big)
\nonumber\\
&~&\qquad +X_A+
\sum_{I}{X_{\overline{N^I}}}+
\mcal{A}_{GGX}^{hidden~sector};~~~~
\end{eqnarray}
\begin{eqnarray}
\label{yxx}      
\mcal{A}_{YXX}&=&-2\Big[{X_{H^\mathcal{U}}}^2-{
      X_{H^\mathcal{D}}}^2
     \nonumber \\
    & & \qquad +
    \sum_{i}\Big({X_{Q^i}}^2-2~{X_{\overline{U^i}}}^2+
    {X_{\overline{D^i}}}^2-{X_{L^i}}^2+{X_{\overline{E^i}}}^2
    \Big)\Big]\cdot Y_L.~~~~
\end{eqnarray}

We have not fixed the normalization of the hypercharges, 
and we used the standard GUT normalization for the 
generators of the non-Abelian gauge groups:
\beq
\mbox{tr}[t_a~t_b]~=~\scr{N}~\delta_{ab},~~~~\mbox{with}~\scr{N}=\mbox{
$\frac{1}{2}$}.
\eeq
In this convention one has
\beq\label{555}
{g_C}^2~k_C~=~{g_W}^2~k_W~=~{g_Y}^2~k_Y~=~{g_X}^2~k_X~=
~2~{g_{s}}^2,
\eeq
$g_s$ being the string coupling  constant; for the factor 
of 2 in 
Eq.~(\ref{555})  and  a discussion of the mismatch between the 
conventions of  GUT  and string amplitudes 
  see Ref.~\cite{cvetic} and 
Ref.~\cite{kaplunovsky}. We assume gauge coupling 
unification within the 
context of string theory, see Ref.~\cite{ginsparg}, 
so phenomenology  requires 
 ${g_C}^2=\frac{10}{3\scr{N}}{Y_L}^2{g_Y}^2$, hence
\beq
k_C=k_W=\frac{3}{5}\cdot\frac{k_Y}{4\cdot{Y_L}^2}~.
\eeq
%

{}Even without knowing the exact values for the $X$-charges,
 one can check whether the Green-Schwarz conditions forbid
Section~\ref{Consr}'s way of achieving $R_p$. \emph{E.g.}
$\mathcal{A}_{CCX}$ must be of the same fractionality as
$\mathcal{A}_{WWX}$: From Eqs.~(\ref{integerdelta},\ref{wwx}) one finds
 (with
$\Delta^{Q,L}_{11}=0$, $\mathcal{N}_g$ is the number of 
generations)
\beq
\mathcal{A}_{WWX}=\frac{1}{2}\Big[X_{H^\mathcal{U}}+
X_{H^\mathcal{D}}+\mathcal{N}_g~(3X_{Q^1}+X_{L^1})+
\sum_{i=1}^{\mathcal{N}_g}\big(3
\Delta^{Q}_{i1}+\Delta^{L}_{i1}\big)   \Big].
\eeq
Rearranging and using  Eqs.~(\ref{cc},\ref{ccx}) gives
\beqn
\lefteqn{
  3X_{Q^1}+X_{L^1}
} \nonumber \\
& &=\frac{1}{\mathcal{N}_g }
\Big[2{\mathcal{A}_{CCX}}-
\big(3(\Delta^Q_{21}+\Delta^Q_{31}+...)+(\Delta^L_{21}+
\Delta^L_{31}+...)+X_{H^\mathcal{U}}+X_{H^\mathcal{D}}\big)
\Big]\nonumber\\
  &&=\frac{1}{\mathcal{N}_g}\Big[\sum_{i=1}^{\mathcal{N}_g}
\big(X_{Q^i}+
X_{H^\mathcal{U}}
+X_{\overline{U^i}}\big)+  \sum_{i=1}^{\mathcal{N}_g}
\big( X_{Q^i}+
X_{H^\mathcal{D}}+X_{\overline{D^i}} \big)  \nonumber\\
 &~&\qquad  -\big(3(\Delta^Q_{21}+\Delta^Q_{31}+...)
+(\Delta^L_{21}+\Delta^L_{31}+...)+(1+\mathcal{N}_g)(X_{H^\mathcal{U}}+
X_{H^\mathcal{D}})\big)\Big]\nonumber\\
&&=\frac{\mbox{integer}}{\mathcal{N}_g}.\label{threeee}
\eeqn
In the following, we work with $\mathcal{N}_g=3$.  One can see that
the  condition above is compatible  with Eq.~(\ref{CASE}). This match is
not given for $B_p$ and $L_p$, see Appendix~\ref{bplp}.

%
%
%
\section{\label{pHeNo}Phenomenological 
Constraints from  Quarks and  Charged Leptons}
\cleqn

In this section, we use phenomenologically acceptable forms of mass
matrices for up-quarks, down-quarks, charged leptons, and the CKM
matrix, and determine the $U(1)_X$ charge assignments consistent with
them.  We make  full use of the anomaly cancellation conditions which were  derived
in the previous section.  There are five viable patterns for quark mass
matrices Eqs.~(\ref{fihr}--\ref{droi}), and we will be left with three
real parameters ($X_{L^1}$, $\Delta_{21}^L$, $\Delta_{31}^L$) for each
pattern, as shown in Table~\ref{Table1}.  At this point, 
the $U(1)_X$ charges for two right-handed neutrinos are left free.  
Combining this with the requirement of automatic $R$-parity
conservation, we arrive at Table~\ref{Table2} where the parameters
$\Delta^L_{31},~\zeta,~\Delta^{\!H},~\nu, ~\Delta^{\overline{N}}_{21}$
are constrained to be integers.  Eventhough
each of the five patterns is phenomenologically viable, we
pick the patterns Eqs.~(\ref{zwo}) and (\ref{oins}) because the CKM
matrix comes out most successfully [the middle one in Eq.~(\ref{6})].

To identify phenomenologically acceptable mass matrices, we follow
Ref.~\cite{dt}. The mass eigenvalues are given at 
the  GUT scale, see Ref.~\cite{rrr,n},\footnote{For fields
except the top-quark the fermion masses renormalize
practically only according to the anomalous dimensions
due to gauge interactions, and hence their 
intergenerational ratios do not renormalize.}
\begin{eqnarray}
m_e:m_\mu:m_\tau&\sim&{\lambda_c}^{4~\mbox{\scriptsize or}~5}:{\lambda_c}^2:
1,\\ 
m_\tau:m_b&\sim& 1,\\
m_d:m_s:m_b&\sim&{\lambda_c}^4:{\lambda_c}^2:1,\\
m_b:m_t&\sim&{\lambda_c}^{0,1,2~\mbox{\scriptsize or}~3}\;\langle H^\mcal
{D}\rangle\big/\langle H^\mcal{U}\rangle,\label{4}\\ 
m_u:m_c:m_t&\sim&
{\lambda_c}^8:{\lambda_c}^4:1,\label{5}\\
m_t&\sim&\langle H^\mcal{U}\rangle,
\end{eqnarray}
and in addition one has the three ans\"atze
\begin{eqnarray}
 {\bsym{V^{\!C\!K\!M}}}
\sim\left(\begin{array}{lll} 1 & 1 & {\lambda_c}^2 \\
1 & 1 & {\lambda_c}^2 \\
{\lambda_c}^2 & {\lambda_c}^2 & 1 \end{array}\right)~
\mbox{or}~\left(\begin{array}{lll} 1 & {\lambda_c} & 
{\lambda_c}^3 \\
{\lambda_c} & 1 & {\lambda_c}^2 \\
{\lambda_c}^3 & {\lambda_c}^2 & 1 \end{array}\right)~
\mbox{or}~\left(\begin{array}{lll}
1 & {\lambda_c}^2 & {\lambda_c}^4 \\
{\lambda_c}^2 & 1 & {\lambda_c}^2 \\
{\lambda_c}^4 &{\lambda_c}^2 & 1 \end{array}\right)\!\!,
\label{6}
\end{eqnarray}
where the coefficients of $\mathcal{O}(1)$ in each component
of these matrices are implicit. 
${\lambda_c}\sim0.22$
 is the Wolfenstein parameter, {\emph{i.e.}} the (sine of the) 
Cabibbo angle,
$\langle H^\mcal{U}\rangle$ and 
$\langle H^\mcal{D}\rangle$ denote the VEVs of the 
two neutral Higgs scalars, ${\bsym{V^{\!C\!K\!M}}}$ is the 
Cabibbo-Kobayashi-Maskawa matrix. 
The first\footnote{This shape of ${\bsym{V^{\!C\!K\!M}}}$ 
  was only recently suggested in  Ref.~\cite{0202101}.} 
and the 
last choice for the CKM matrix require accidental cancellations of
${\cal O}(\epsilon)$ among the unknown $\mathcal{O}(1)$-coefficients.
The second choice is slightly preferred, which is
why we will eventually discard the first and the third choice.
 
If one is dealing with $U(1)_X$ and one flavon superfield,  the only
pairs of $u$- and $d$-type quark mass  matrices (after the K\"ahler
potential has been diagonalized and thus textures have been filled
up)  which can be generated \'a la FN
and which simultaneously reproduce the quark masses  and mixings as 
displayed above  are (see Refs.~\cite{blr,cl,haba,dt}; 
the textures in Eq.~(\ref{funef}) are presented for the 
first time)  
\begin{eqnarray}
\label{fihr}
&{\boldsymbol{G^{(U)}}}~\sim~\left(\begin{array}{lll}
{\lambda_c}^8 & {\lambda_c}^{6\phantom 3} & {\lambda_c}^4
 \\
{\lambda_c}^{14} & {\lambda_c}^4 & {\lambda_c}^2 \\
{\lambda_c}^{12} & {\lambda_c}^2 & 1 \end{array}\right),
~~~&{\boldsymbol{G^{(D)}}}
~\sim~{\lambda_c}^{x}~
\left(\begin{array}{lll}
{\lambda_c}^{4\phantom 3} & {\lambda_c}^{4\phantom 3} & 
{\lambda_c}^4 \\
{{\lambda_c}}^{10} & {{\lambda_c}}^2 & {{\lambda_c}}^2 \\
{{\lambda_c}}^8 & 1 & 1 \end{array}\right), \\
\label{zwo}
&{\boldsymbol{G^{(U)}}}~\sim~\left(\begin{array}{lll}
{{\lambda_c}}^8 & {{\lambda_c}}^{5\phantom 3} & 
{{\lambda_c}}^3 \\
{{\lambda_c}}^{13} & {{\lambda_c}}^4 & {{\lambda_c}}^2 \\
{{\lambda_c}}^{11} & {{\lambda_c}}^2 & 1 \end{array}
\right),~~~&
{\boldsymbol{G^{(D)}}}
~\sim~{{\lambda_c}}^{{x}}~
\left(\begin{array}{lll}
{{\lambda_c}}^{4\phantom 3} & {{\lambda_c}}^{3\phantom 3}
 & {{\lambda_c}}^3 \\
{{\lambda_c}}^9 & {{\lambda_c}}^2 & {{\lambda_c}}^2 \\
{{\lambda_c}}^7 & 1 & 1 \end{array}\right),\\
\label{funef}
&\boldsymbol{G^{(U)}}~\sim~
\left(\begin{array}{lll}
{{\lambda_c}}^8 & {{\lambda_c}}^{4\phantom 3} & 
{{\lambda_c}}^{2} \\
{{\lambda_c}}^{8\phantom 3} & {{\lambda_c}}^4 & 
{{\lambda_c}}^2 \\
{{\lambda_c}}^6 & {{\lambda_c}}^2 & 1 \end{array}\right),
~~~&
{\boldsymbol{G^{(D)}}}~\sim~{{\lambda_c}}^{{x}}~
\left(\begin{array}{lll}
{{\lambda_c}}^{4\phantom 3} & {{\lambda_c}}^{2\phantom 3}
 & {{\lambda_c}}^2 \\
{{\lambda_c}}^4 & {{\lambda_c}}^2 & {{\lambda_c}}^2 \\
{{\lambda_c}}^2 & 1 & 1 \end{array}\right),\\
\label{oins}
&\boldsymbol{G^{(U)}}~\sim~
\left(\begin{array}{lll}
{{\lambda_c}}^8 & {{\lambda_c}}^{5\phantom 3} &
 {{\lambda_c}}^{3} \\
{{\lambda_c}}^{7\phantom 3} & {{\lambda_c}}^4 &
 {{\lambda_c}}^2 \\
{{\lambda_c}}^5 & {{\lambda_c}}^2 & 1 \end{array}\right),
~~~&
{\boldsymbol{G^{(D)}}}~\sim~{{\lambda_c}}^{{x}}~
\left(\begin{array}{lll}
{{\lambda_c}}^{4\phantom 3} & {{\lambda_c}}^{3\phantom 3}
 & {{\lambda_c}}^3 \\
{{\lambda_c}}^3 & {{\lambda_c}}^2 & {{\lambda_c}}^2 \\
{{\lambda_c}} & 1 & 1 \end{array}\right),\\
\label{droi}
&{\boldsymbol{G^{(U)}}}~\sim~\left(\begin{array}{lll}
{{\lambda_c}}^8 & {{\lambda_c}}^{6\phantom 3} & 
{{\lambda_c}}^4 \\
{{\lambda_c}}^{6\phantom 3} & {{\lambda_c}}^4 & 
{{\lambda_c}}^2 \\
{{\lambda_c}}^4 & {{\lambda_c}}^2 & 1 \end{array}\right),
~~~&
{\boldsymbol{G^{(D)}}}
~\sim~{{\lambda_c}}^{{x}}~
\left(\begin{array}{lll}
{{\lambda_c}}^{4\phantom 3} & {{\lambda_c}}^{4\phantom 3} 
& {{\lambda_c}}^4 \\
{{\lambda_c}}^2 & {{\lambda_c}}^2 & {{\lambda_c}}^2 \\
1 & 1 & 1 \end{array}\right).~~~~~~~
\end{eqnarray}
Here $x=0,1,2,3$, except for Eq.~(\ref{zwo}), where the choice is
limited to $x=0,1,2$. 
The first and the last of these pairs of matrices lead to 
the third choice for the CKM matrix in Eq.~(\ref{6}), the 
third pair corresponds to the first choice in 
Eq.~(\ref{6}). 
The second pair does not give 
$m_b:m_t\sim{\lambda_c}^{3}\langle H^\mcal{D}\rangle
\big/\langle H^\mcal{U}\rangle$,
see below Eq.~(\ref{gfgfgfg}). As a spot check, we 
investigated  the validity of ${\boldsymbol{G^{(U)}}}$ 
in Eq.~(\ref{oins})  with an ensemble of 3000 
\emph{Mathematica}$^\copyright$-randomly generated sets 
of $\mathcal{O}(1)$ and complex ${g^{(U)}}_{ij}$. In 
Figure~1 the  logarithm to base $\lambda_c$ 
 of the positive
   square roots of the eigenvalues of 
${\boldsymbol{G^{(U)}}}{\boldsymbol{G^{(U)}}}^\dagger$
is plotted against the 3000 trials. The result  
agrees  well with Eq.~(\ref{5}), apart from 
largish scatters. 
\begin{figure}
\centering 
\includegraphics[width=0.6\columnwidth]{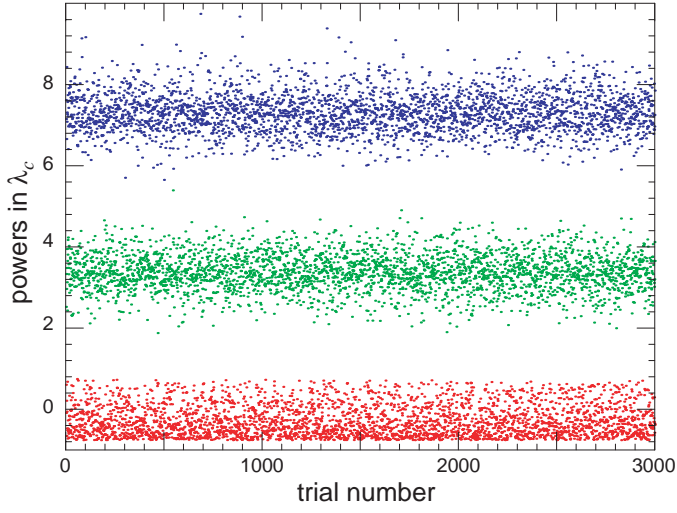}
%
%
%
%
%
%
\caption{The powers in $\lambda_c$ of the positive square 
  roots of the three eigenvalues of ${\boldsymbol{G^{(U)}}}
  {\boldsymbol{G^{(U)}}}^\dagger$ for an ensemble of 3000
  \emph{Mathematica}$^\copyright$-randomly generated sets of
  ${g^{(U)}}_{ij}$ which are complex and of $\mathcal{O}(1)$.}
\label{fig:Statistical-Test}
\end{figure}

In order to reproduce the patterns 
Eqs.~(\ref{fihr}-\ref{droi}),
 the $X$-charges have to fulfill, see Ref.~\cite{dt},
\begin{eqnarray}\label{rrr}
X_{Q^i}+X_{H^\mcal{U}}+X_{\overline{U^j}}&=&\left
(\begin{array}{rrr}
8 & 5+y  &  3+y        \\
7-y &4  & 2           \\
5-y & 2  & 0     \end{array}\right)_{\!\!ij}~~,\\
\nonum\\
\label{RRR}X_{Q^i}+X_{H^\mathcal{D}}+X_{\overline{D^j}}
&=&\left(
\begin{array}{rrr}
4+x & 3+y+x  &  3+y+x        \\
3-y+x &2+x  & 2+x             \\
1-y+x & x & x      \end{array}\right)_{\!\!ij}, \\
X_{L^i}+X_{H^\mathcal{D}}+X_{\overline{E^i}}&=&
\left(\begin{array}{r}
4+z+x\\
2+x \\
x \end{array}\right)_{\!\!i}\;,
\end{eqnarray}
with
\begin{eqnarray}\label{gfgfgfg}
x=0,1,2,3\quad {\rm if}\quad y=-7,-1,0,1\quad~~ {\rm or}
~~ \quad x=0,1,2\quad 
{\rm if} \quad y=-6\nonumber\\ 
\mbox{and}~~~~~~~  \label{xyz} z=0,1.
~~~~~~~~~~~~~~~~~~~~~~~~~~~~~~
\end{eqnarray}
Eqs.~(\ref{fihr}-\ref{droi}) are in order of increasing 
$y$. The cases $y=-6,-7$ necessarily need
supersymmetric zeros in the (1,2)- and (1,3)-entries
of $\bsym{G^{(U)}},\bsym{G^{(D)}}$, which is why
$y=-6$, $x=3$ has to be excluded. $x=3$ is our preferred 
value, since it requires a small 
value of
$\tan\beta$, \emph{i.e.} $\langle H^\mcal{U}\rangle$  is of the 
same order of magnitude as $\langle H^\mcal{D}\rangle$,
which we find  more natural than $\langle H^\mathcal{U}
\rangle\gg\langle H^\mathcal{D}\rangle$. In the rest of 
this text we shall not deal with $y=-7,-1,1$ anymore, 
because $y=-6,0$ produces the best fit to the CKM matrix,
 not 
requiring any (mild) fine-tuning [as has  already 
been stated
below Eq.~(\ref{6})]. The set of $X$-charges 
which is 
constrained by the conditions of anomaly cancellation
[$S\!M$ with $U(1)_X$] in  Section~\ref{anom}
  and which gives rise to the 
phenomenology  explained above is displayed in 
Table~\ref{Table1}, 
see Ref.~\cite{dt}.

There is however an important  no-go
\emph{caveat}.  Ref.~\cite{Espinosa:2004ya} shows that if  matrices 
with supersymmetric zeros predict  a CKM matrix which is in gross
disagreement with the experimentally measured CKM matrix, then this
persists even when the supersymmetric zeros  are filled in, although
``on first sight"  the matrices in Eqs.~(\ref{fihr},\ref{zwo}) produce
nice results.  
The K\"ahler potential for the left-handed quarks affects both the
$u$- and the $d$-type mass matrices,  hence the  entries in the
physical mass matrices are very correlated after the
canonicalization. This is why  during diagonalizing  different 
terms can cancel. See also Refs.~\cite{kprvv,jj2}.  Therefore the choices
$y=-6$ and $-7$ actually do not reproduce the matrices in
Eqs.~(\ref{fihr},\ref{zwo}) and hence phenomenologically
uninteresting.  Nonetheless, we will discuss these cases in the paper,
because it is interesting to see that they are excluded by other
reasons.

Should one wish to impose $SU(5)$ invariance on the $U(1)_X$ charge
assignments, one has to work with $y=1$ and
$z=\Delta^L_{21}=\Delta^L_{31}=0$.  We will not be able to do so,
however, as we  see below, but  this is consistent with 
our philosophy of not having an
additional mass scale for grand unification. The $X$-charges in Table~\ref{Table1} are not compatible with  invariance under flipped $SU(5)\times U(1)^\prime$.

\begin{table}
\begin{center}
\vspace{-1cm}
\begin{tabular}{|rcl|}
\hline
$~\phantom{\Big|}X_{H^\mathcal{D}}$
 &$=$&$\frac{1}{ 54 + 9x   + 6z}~\big[-18 + 36x+ 18y $\\
 &$~$&$\qquad+6x^2+2z^2+5xz-X_{L^1}\left(36+6x+9z\right)$\\
 &$~$&$\qquad-\left(12+2x+2z\right) 
\Delta^L_{21}-\left(6+2x+2z\right)
\Delta_{31}^L~\big]$\\
$~\phantom{\Big|}X_{H^\mathcal{U}}$&$=$&$-
X_{H^\mathcal{D}}-z $\\
$\phantom{\Big|}X_{Q^1}$&$=$&$\frac{1}{9}\big[30 - 
3X_{L^1} - \Delta^L_{21}-\Delta^L_{31} + 3x + 6y 
+ 4z\big]$\\
$~\phantom{\Big|}X_{Q^2}$&$=$&$X_{Q^1}-1-y $\\
$~\phantom{\Big|}X_{Q^3}$&$=$&$X_{Q^1}-3-y $\\
$~\phantom{\Big|}X_{\overline{D^1}}$&$=$&$-X_{
H^\mathcal{D}}-X_{Q^1}+4+x $\\
$~\phantom{\Big|}X_{\overline{D^2}}$&$=$&$X_{
\overline{D^1}}-1+y $\\
$~\phantom{\Big|}X_{\overline{D^3}}$&$=$&$X_{
\overline{D^1}}-1+y $\\
$~\phantom{\Big|}X_{\overline{U^1}}$&$=$&$X_{
H^\mathcal{D}}-X_{Q^1}+8+z $\\
$~\phantom{\Big|}X_{\overline{U^2}}$&$=$&$X_{
\overline{U^1}}-3+y $\\
$~\phantom{\Big|}X_{\overline{U^3}}$&$=$&$X_{
\overline{U^1}}-5+y $\\
$~\phantom{\Big|}X_{L^2}$&$=$&$X_{L^1}+\Delta^L_{21}$\\
$~\phantom{\Big|}X_{L^3}$&$=$&$X_{L^1}+\Delta^L_{31}$\\
$~\phantom{\Big|}X_{\overline{E^1}}$&$=$&$-X_{
H^\mathcal{D}}        {+~4} - X_{L^1} + x + z $\\
$~\phantom{\Big|}X_{\overline{E^2}}$&$=$&$-X_{
H^\mathcal{D}}  {+~2} - X_{L^1} + x -\Delta^L_{21}      
$ \\ 
$~\phantom{\Big|}X_{\overline{E^3}}$&$=$&$-X_{
H^\mathcal{D}}  \phantom{+~4} - X_{L^1} + x -
\Delta^L_{31}             $ \\  & &\\ \hline
\end{tabular}
\caption{\label{Table1} The constrained 
  $X$-charges to reproduce phenomenologically acceptable mass matrices
  Eqs.~(\ref{fihr}-\ref{droi}), with the normalization $X_A=-1$.
  $X_{L^1},~\Delta^L_{21},~\Delta^L_{31}$ are real numbers, for
  $x,y,z$ see Eq.~(\ref{xyz}).  $SU(5)$ invariance would require $y=1$,
  $z=\Delta^L_{21}=\Delta^L_{31}=0$.}
\end{center}
\end{table}

We now check whether Table~\ref{Table1} can be 
combined with 
$R_p$ being conserved by virtue of the $X$-charges.
 Eq.~(\ref{integerdelta})   is fulfilled  
if the  $\Delta^L_{...}$ are integer.  
Eq.~(\ref{trilin}) is 
automatically fulfilled, as seen from 
Eqs.~(\ref{rrr},\ref{RRR}). 
With Table~\ref{Table1} we see 
\beq
3X_{Q^1}+X_{L^1}=10-\mbox{$\frac{1}{3}$}(\Delta^L_{21}
+\Delta^L_{31})+x+2y+\mbox{$\frac{4}{3}$}z.
\eeq
Thus $3X_{Q^1}+X_{L^1}$ is integer if and only if, 
now working with the $\Delta^L_{...}$ being integers,
\beq\label{ddzz}
\Delta^L_{21}+\Delta^L_{31}=3\zeta+z,
\eeq
and $\zeta$ is integer. With this constraint,
 Eq.~(\ref{l0.5}) is  
 fulfilled for  a special choice of $X_{H^\mathcal{D}}$,
for which we introduce the integer parameter $\Delta^{\!H}=X_{L^1}-X_{H^\mathcal{D}}-\frac{1}{2}$. 
So the union of Table~\ref{Table1} and 
conserved $R_p$ is indeed 
possible as given in  Table~\ref{Table2}. 
Note that all  
conclusions so far are applicable to 
the case of any number of right-handed neutrinos, in 
Table~\ref{Table2} however we have restricted 
ourselves to two 
right-handed neutrinos.
\begin{table}
\begin{center}
\vspace{-1cm}
\begin{tabular}{|rcl|}
\hline
  & & \\
$~~~\phantom{\Big|}X_{H^\mathcal{D}}$
 &$=$&$\frac{1}{10~(6+x+z )}~\Big(12y + 2x ~(2x + 11 + z 
- 2\Delta^{\!H}) $\\
       & &$  - z~(11 + 6 \Delta^{\!H}) - 
  4 ~(6 + 6 \Delta^{\!H} - \Delta^L_{31})-4~(6+x+z)\zeta
\Big)~~~$\\
$~\phantom{\Big|}X_{H^\mathcal{U}}$
 &$=$&$-z-\phantom{\Big|}X_{H^\mathcal{D}}  $\\
$~\phantom{\Big|}X_{Q^1}$&$=$&$\frac{1}{3}\Big(
      \frac{19}{2} - X_{H^\mathcal{D}} + x + 2y + z - 
\Delta^{\!H} - \zeta\Big)$\\
$~\phantom{\Big|}X_{Q^2}$&$=$&$X_{Q^1}-1-y $\\
$~\phantom{\Big|}X_{Q^3}$&$=$&$X_{Q^1}-3-y $\\
$~\phantom{\Big|}X_{\overline{D^1}}$&$=$&$-X_{
H^\mathcal{D}}-X_{Q^1}+4+x $\\
$~\phantom{\Big|}X_{\overline{D^2}}$&$=$&$X_{
\overline{D^1}}-1+y $\\
$~\phantom{\Big|}X_{\overline{D^3}}$&$=$&$
X_{\overline{D^1}}-1+y $\\
$~\phantom{\Big|}X_{\overline{U^1}}$&$=$&$
X_{H^\mathcal{D}}-X_{Q^1}+8+z $\\
$~\phantom{\Big|}X_{\overline{U^2}}$&$=$&$
X_{\overline{U^1}}-3+y $\\
$~\phantom{\Big|}X_{\overline{U^3}}$&$=$&$
X_{\overline{U^1}}-5+y $\\
$~\phantom{\Big|}X_{L^1}$&$=$&$\frac{1}{2}+
X_{H^\mathcal{D}}+\Delta^{\!H}$\\
$~\phantom{\Big|}X_{L^2}$&$=$&$X_{L^1} + z-
\Delta^L_{31}+  3\zeta $\\
$~\phantom{\Big|}X_{L^3}$&$=$&$X_{L^1}\phantom{+z}~~
+\Delta^L_{31}$\\
$~\phantom{\Big|}X_{\overline{E^1}}$&$=$&$-X_{
H^\mathcal{D}}        {+~4} - X_{L^1} + x + z $\\
$~\phantom{\Big|}X_{\overline{E^2}}$&$=$&$-X_{
H^\mathcal{D}}  {+~2} - X_{L^1} + x- z +
\Delta^L_{31} - 3\zeta      $ \\ 
$~\phantom{\Big|}X_{\overline{E^3}}$&$=$&$-
X_{H^\mathcal{D}}  \phantom{+~4} - X_{L^1} + 
x \phantom{-z}~~-\Delta^L_{31}             $ \\ 
$~\phantom{\Big|}X_{\overline{N^{     1}}}     
$&$=$&$ \frac{1}{2}+\nu   $ \\ 
$~\phantom{\Big|}X_{\overline{N^{2}}}     $&$=$&$
 \frac{1}{2}+\nu+\Delta^{\overline{N}}_{21} $ \\
  & &\\ \hline
\end{tabular}
\caption{\label{Table2} 
  In addition to Table~\ref{Table1}, the automatic $R_p$ conservation
  was imposed.  $\Delta^L_{31},~\zeta,~\Delta^{\!H},~\nu,
  ~\Delta^{\overline{N}}_{21} $ are integers, for $x,y,z$ see
  Eq.~(\ref{xyz}). $SU(5)$ invariance would require $y=1$,
  $z=\Delta^L_{31}=\zeta=0$. We have restricted ourselves to two
$\overline{N^I}$.}
\end{center}
\end{table}

For the upcoming calculations it is useful to know that 
%
\beqn\label{59999}
\lefteqn{
  X_{L^i}+X_{H^\mathcal{U}}+X_{\overline{E^j}}
} \nonumber \\
  &=&x+\left(\begin{array}{ccc}  4+z & 2  & 0\\  4+z & 2  
      & 0\\  4+z & 2  & 0\\ \end{array}\right)_{ij}+\left(
    \begin{array}{ccc}  0 & -\Delta^L_{21}  & -\Delta^L_{31}
      \\\Delta^L_{21}  & 0  &\Delta^L_{21}-\Delta^L_{31} \\
      \Delta^L_{31}    &\Delta^L_{31}-\Delta^L_{21}   & 0\\ 
    \end{array}\right)_{ij}  \nonumber \\
&=& x+
\left(\begin{array}{ccc}  4+z & 2  & 0\\  4+z & 2  & 0\\ 
    4+z & 2  & 0\\ \end{array}\right)_{ij} \nonumber \\
& &\qquad 
 +\left(
  \begin{array}{ccc}  0 & -z+\Delta^L_{31}-3\zeta  & -
    \Delta^L_{31} \\ z - \Delta^L_{31} + 3 \zeta  & 0  & 
    z - 2 \Delta^L_{31} + 3 \zeta \\\Delta^L_{31}    &-z 
    + 2 \Delta^L_{31} - 3 \zeta  & 0\\ \end{array}
\right)_{ij}.
\eeqn

{It is worth pointing  out that there already exists a 
model in the literature which fulfills all the necessary 
constraints for $R_p$ being conserved due to the 
$X$-charges, namely Ref.~\cite{eir} (with however three 
generations of right-handed neutrinos). 
This model 
is in the tradition of the papers Ref.~\cite{9612442} 
 and 
Ref.~\cite{98}. In the former, 
a general analysis of  $D$-flat directions and 
the seesaw  mechanism leads  to conserved
 $R_p$, in the latter, the authors worked out a 
concrete model. They considered 
three beyond-$S\!M$ $U(1)$s, two of them being 
generation-dependent and non-anomalous, one being 
generation-independent and anomalous. In   
Ref.~\cite{eir} the 
before mentioned  symmetries were not gauged separately 
but together, so that this model falls into the category
 considered here;  in our notation, the authors
 work with 
$x=3$, $y=0$, $z=0$, $\Delta^L_{31}=-3$, $\zeta=-2$,
$\Delta^{\!H}=1$.

}

%
%
\section{\label{vEv}The
 VEV of the Flavon; Tadpoles}
\cleqn

Because we would like to construct a complete theory of flavor out of
only two mass scales, $M_{grav}$ and $m_{3/2}$, the mass scale of the
$U(1)_X$ breaking must be a {\it derived}\/ scale.  Indeed, the vacuum
expectation value of the flavon is determined dynamically thanks to
the anomalous nature of $U(1)_X$.  We show explicitly that our $X$-charge
assignments can successfully lead to an expansion parameter $\epsilon
= \langle A\rangle/M_{grav} = 0.171$--0.221$\simeq \lambda_c$ as
desired phenomenologically.  We, however, point out an important
\emph{caveat} in a class of string-derived models.  We also show that
tadpoles are of no concern.

In the string-embedded FN framework the expansion parameter
 $\epsilon$ (which will be identified with $\lambda_c$)
 has its origin solely in the 
Dine-Seiberg-Wen-Witten mechanism, due to which the 
coefficient of the Fayet-Iliopoulos term is radiatively 
generated. One has, see Ref.~\cite{cvetic}, 
\beq 
 \xi_X~=~{g_s}^2~\frac{\mathcal{A}_{GGX}}{192\pi^2}~
{M_{grav}}^2
\eeq
(${\xi_X}^{tree~level}$ is zero in local 
supersymmetry, see Ref.~\cite{barbieri}). This gives
\beq\label{aaa}
 \langle A \rangle~ =~\sqrt{-\frac{\xi_X}{X_A}~},
\eeq
supposing that no other fields break $U(1)_X$. With 
$X_A=-1$, using Eq.~(\ref{cc}) to eliminate 
$\mathcal{A}_{GGX}$ in 
favor of $\mathcal{A}_{CCX}$, Eq.~(\ref{ccx}), 
Eq.~(\ref{555}) 
and Table~\ref{Table1} one finds
\begin{equation}\label{wewofaei}
\langle A \rangle~=~\frac{g_C}{4\pi\sqrt{2}}
\sqrt{3(6+x+z)~}\cdot M_{grav}.
\eeq
Similar calculations with 
similar results have been presented in Refs.~\cite{98,buk}.
  Using Eq.~(\ref{eeppss}), replacing
(see Section~\ref{scenario})
\beq\label{wca}
M=M_{grav}
\eeq 
and evaluating   
$g_C[M_{GUT}=2.2 \times10^{16}\mbox{~GeV}]
\approx0.72$ we obtain
\beq
0.171\leq \epsilon \leq 0.221.
\eeq
So $\eps=\lambda_c$; the best match is
obtained  for $x=3$, $z=1$.\footnote{For $x=z=0$ 
one has that 
$m_u\sim\eps^8~m_t$ is a factor of seven below the 
desired value.} 

\emph{However, there 
is a very important \emph{caveat} which one
 should keep in mind:}  Eq.~(\ref{wca}) together with  the 
 assumption that
 the dimensionless prefactors like ${g^{(...)}}_{ij}$ 
are of $\mathcal{O}(1)$ might well not be justified by 
superstring theory. In Ref.~\cite{cvetic} it is nicely and 
clearly demonstrated   how a prototype string theory 
(Ref.~\cite{prototype}) produces an effective 
supersymmetric 
theory with a superpotential,  including the 
coupling constants. Translating their result to our 
notation we get \emph{e.g.} instead of  
Eq.~(\ref{huns})
\beqn\label{sszig}
&&\Theta\big[X_{Q^i}+X_{H^{\mcal{U}}}+
X_{\overline{U^j}}\big]~\cdot~\Omega\big[X_{Q^i}+
X_{H^{\mathcal{U}}}+
X_{\overline{U^j}}\big]~~\times~{g}_{C}~~
\sqrt{\frac{k_C~}{2}~}~~\mathtt{C}_{{X_{Q^i}+
X_{H^\mathcal{U}}+X_{\overline{U^j}}}}\nonumber\\&&
~~~~~\times~\mathtt{I}_{{X_{Q^i}+X_{H^\mathcal{U}}+
X_{\overline{U^j}}}}~~\bigg(\frac{A}{\pi~M_{grav}}
\bigg)^{{X_{Q^i}+X_{H^\mathcal{U}}+X_{\overline{U^j}}}}.
\eeqn
$\mathtt{C}_{...}$ is a $\mathcal{O}(1)$ Clebsch-Gordan 
coefficient and  $\mathtt{I}_{...}$ a world sheet 
integral. For large ${X_{Q^i}+X_{H^\mathcal{U}}+
X_{\overline{U^j}}}$ naively one would expect 
$\mathtt{I}_{{X_{Q^i}+X_{H^\mathcal{U}}+
X_{\overline{U^j}}}}\sim {\mathtt{I}_{1}}^{{X_{Q^i}+
X_{H^\mathcal{U}}+X_{\overline{U^j}}}}$ (with  
${\mathtt{I}_{1}}\sim70$), but due to destructive 
interference effects of the integrands  
actually\footnote{We thank  
Mirjam Cveti{\v{c}} for pointing 
this out.} 
\beq\label{bsjawohl}
\mathtt{I}_{{X_{Q^i}+X_{H^\mathcal{U}}+
X_{\overline{U^j}}}}\ll {70}^{{X_{Q^i}+
X_{H^\mathcal{U}}+X_{\overline{U^j}}}},
\eeq
and therefore
\beq\label{siebzig}
{G^{(U)}}_{\!ij}~\ll~\Theta[X_{Q^i}+...]\cdot 
\Omega[X_{Q^i}+...]\cdot\mathtt{C}_{X_{Q^i}+...}~~
g_C~\sqrt{\frac{k_C~}{2}}~~\bigg(\frac{70~\lambda_c}{\pi}
\bigg)^{{X_{Q^i}+X_{H^\mathcal{U}}+X_{\overline{U^j}}}}. 
\eeq
This does not necessarily guarantee $\epsilon$ and the 
dimensionless prefactors to have  the desired values.
In this paper we simply assume that this is nevertheless
the case. For another
discussion on the calculation of fermionic mass terms in 
string derived models see \emph{e.g.} Ref.~\cite{ALON}.

Below the  $U(1)_X$ breaking scale 
$\epsilon\cdot M_{grav}$ there are three singlets 
$\{A^\prime, \overline{N^I}\}$, with 
$A^\prime=A-\langle A \rangle$.
One must thus wonder whether these lead to tadpoles
causing  quadratic divergences and thus possibly
destabilizing the hierarchy between the weak scale and
 $M_{grav}$, see Refs.~\cite{tp1,tp2,tp3,tp4}. 
First of all, in our model $R_p$
is conserved before and after the breaking of $U(1)_X$. 
This prevents any $\overline{N^I}$-tadpole term.
Second,  $A^\prime$-tadpoles are harmless,
due to the high mass of  $A^\prime$, given by
 \mbox{$\eps\cdot M_{grav}$.}

%
%
%

\section{\label{uP} $\boldsymbol{\mu}$-Parameter and Proton Decay}
\cleqn

So far, we are left with the two patterns 
Eqs.~(\ref{zwo},\ref{oins}) 
with $y=-6,0$, respectively, with possible 
choices $x=0,1,2,3~\big/~x=0,1,2$, respectively, and $~z=0,1$.  
In this section, we narrow down
the choices further.  First,  the $\mu$-term 
is phenomenologically required to be comparable to $m_{3/2}$.  This selects
$z=1$.  Another requirement is the adequate stability of the proton
against Planck scale $D=5$ operators, which eliminates $y=-6$ and
prefers larger $x$.  The resulting $U(1)_X$ charge assignments are
shown in Table~\ref{Table2.5}.

In order to get a satisfactory $\mu$-term we have to rely on the GM
mechanism for $\mu\propto m_{3/2}$, since
$X_{H^\mathcal{U}}+X_{H^\mathcal{D}}=24$ is not possible, see
Table~\ref{Table1}. This requires
$X_{H^\mathcal{U}}+X_{H^\mathcal{D}}<0$, see Eq.~(\ref{mue}),  hence we
need $z=1.$
Thus, see Eq.~(\ref{wewofaei}),
\beq\label{eps}
\epsilon~=~0.186,~0.198,~0.210,~0.221~~~
\mbox{for}~~~x=0,~1,~2,~3,~~\mbox{respectively}.
\eeq

Next, we consider the proton decay constraints.
We might be  forced not to work with small $x$ and/or 
$y=-6$. This is because of the $R_p$-conserving but nevertheless
proton destabilizing operators 
$\frac{\chi_{ijkl}}{M_{grav}}~Q^iQ^jQ^{k}L^l$
 ($i,j,k$ must not all be the same),
where
\beq\label{siebenzwo} \chi_{ijkl}=\mathcal{O}(1)\times
\eps^{X_{Q^i}+X_{Q^j}+X_{Q^k}+X_{L^l}}
\eeq 
are  dimensionless coupling constants; 
$\overline{U} \overline{U} \overline{D} \overline{E}$
will be dealt with later in this section as well as in 
the second half of Subsection~\ref{subsec:mixing}. 
With Table~\ref{Table2}
one finds that
$Q^1Q^1Q^2L^i$ has the $X$-charges
\beqn\label{qqql}
9+x+y+z+\left(\begin{array}{c}-\zeta\\z-\Delta^L_{31}+
2\zeta\\ \Delta^L_{31}- \zeta\end{array}\right)_{\!\!i}
\eeqn
and
$Q^2Q^1Q^2L^i$ has the $X$-charges
\beqn\label{qqql2}
~~~~~~~~~~~~~~~~~~8+x+z+\left(\begin{array}{c}-\zeta\\
z-\Delta^L_{31}+2\zeta\\ \Delta^L_{31}- \zeta
\end{array}\right)_{\!\!i}.
\eeqn
In comparison, operators involving third-generation 
quarks are  enhanced due to lower $X$-charges.
But their contributions to proton decay are suppressed
 by the entries of the matrices that transform
from the weak base into the mass base, see 
Ref.~\cite{kamu}.

For both equations above one sees that suppressing 
one of the three operators  by choosing an appropriate 
$\zeta$ and/or $\Delta^L_{31}$  makes one or both of the 
others less suppressed. The ``average $X$-charges'' of 
$Q^1Q^1Q^2L^i$ ($\sum_{i}X_{Q^1Q^1Q^2L^i}/3$) and  
$Q^2Q^1Q^2L^i$ are $9+x+y+\frac{4z}{3}$
 and $8+x+\frac{4z}{3}$,  which are not  very high
[note that  already in Ref.~\cite{ky} it was anticipated
that  \emph{e.g.} $z=1$ (our notation) gives a more stable 
proton than 
$z=0$]. 
Thus  already now we
 can see that the model could get into trouble due to 
proton decay if we work  with the wrong choices for 
$x,y,\zeta,\Delta^L_{31}$. For a first crude 
 estimate we use
\beq
\chi~\leq~\frac{\frac{ M_{grav}}{\phantom{\big|}\mbox{{\small{1 GeV}}}}}
   {\sqrt{~2\times10^{17\pm0.7}
\times\frac{\tau}{\mbox{\small years}}~}}~
\times~\frac{m_{squark}}{\mbox{1 TeV}} .
\eeq
$\tau$ is the upper bound on the proton lifetime,
 about $5\times10^{33}~$years for the  
$p\rightarrow\pi^0+\overline{e}$ mode,\footnote{We believe there is a
  typo in Ref.~\cite{prot2} which quotes the limit of $5\times
  10^{32}$~years.}  see \emph{e.g.}
Refs.~\cite{prot1,prot2}.
 The coefficient $2\times10^{17\pm0.7}$ is extracted from 
Ref.~\cite{weinsusy}. 
Being as strict as possible one finds
\beq\label{enaffsapp0}
\chi~\leq~3 \times 10^{-8}~\times~
\frac{m_{squark}}{\mbox{{1 TeV}}}.
\eeq
Now $3\times 10^{-8}\sim0.22^{11}$.  Thus comparing 
the exponent  with the ``average $X$-charges'' it becomes 
apparent that $y=-6$ is not an option, we are thus left 
with $y=0$.\footnote{The choice $y=-6$ was excluded also based on the
consideration in section \ref{pHeNo}.} Furthermore,  one sees that $x=3$ 
(our preferred value; not possible with $y=-6$)
is the safest choice to make, but we have not fixed 
the parameters
of our model enough yet to say that $x=0,1,2$ are not 
viable. 

For a more quantitative investigation we will in the 
next 
Section
rely  on Ref.~\cite{ky}'s treatment of the so called
 ``best-fit''
scenario, taking into account both $QQQL$ and 
$\overline{U}\overline{U}\overline{D}\overline{E}$
(note that $Q Q Q H^\mathcal{D}$
violates $R_p$ and thus is forbidden by the $X$-charges). 
Translated to the  notation of Table~\ref{Table2}, they state that 
the $X$-charges have to fulfill 
(with $m_{squark}=$1~TeV, $y=0$,
$z=1$ )
\beqn  
\mathcal{O}(1)\times\eps^{x+10+\Delta^L_{31} - \zeta}&<&4\times 10^{-8},\nonumber\\
\mathcal{O}(1)\times\eps^{x+10-\Delta^L_{31} + \zeta}~
\frac{\langle H^\mathcal{U}\rangle}{\langle 
H^\mathcal{D}\rangle}&<&1\times 10^{-7},
\eeqn
in order not to be in conflict with experiment. 
With $m_t/m_b$ at high energies being $\sim100$ 
(see Ref.~\cite{ky}) we get from $m_b\sim
\langle H^\mathcal{D}\rangle \eps^x$, $m_t\sim
\langle H^\mathcal{U}\rangle$ that 
$100~\eps^x\sim {\langle H^\mathcal{U}\rangle}/
{\langle H^\mathcal{D}\rangle}$. Thus
\beqn \label{kackizacki} 
\mathcal{O}(1)\times\eps^{x+10+\Delta^L_{31} - 
\zeta}&<&4\times 10^{-8},\nonumber\\
\mathcal{O}(1)\times\eps^{2x+10-\Delta^L_{31} + 
\zeta}&<&1\times 10^{-9}.
\eeqn
Note that our model with $y=0$, $z=1$, 
$\Delta^L_{21}=\Delta^L_{31}=-1$, and $\zeta=-1$ is a special case of
the ``best fit'' model in Ref.~\cite{ky} with $m=1$ in their notation,
while they took $X_{L^3}$ as a free parameter, because they do not impose
the anomaly cancellation conditions nor conserved $R_p$ as a
consequence of the $U(1)_X$ symmetry. A more thorough study  of
proton decay due to higher-dimensional operators and  
$R_p$-conserving $X$-charges will be presented in Ref.~\cite{hlmt}. 

\begin{table}
\begin{center}
\vspace{-1cm}
\begin{tabular}{|rcl|}
\hline
  & & \\
$~~~\phantom{\Big|}X_{H^\mathcal{D}}$
 &$=$&$\frac{1}{10~(7+x)}~\Big(2x ~(2x + 12 
- 2\Delta^{\!H}) $\\
       & &$  - (11 + 6 \Delta^{\!H}) - 
  4 ~(6 + 6 \Delta^{\!H} - \Delta^L_{31})-4~(7+x)\zeta
\Big)~~~$\\
$~\phantom{\Big|}X_{H^\mathcal{U}}$
 &$=$&$-1-\phantom{\Big|}X_{H^\mathcal{D}}  $\\
$~\phantom{\Big|}X_{Q^1}$&$=$&$\frac{1}{3}\Big(
      \frac{21}{2} - X_{H^\mathcal{D}} + x - 
\Delta^{\!H} - \zeta\Big)$\\
$~\phantom{\Big|}X_{Q^2}$&$=$&$X_{Q^1}-1 $\\
$~\phantom{\Big|}X_{Q^3}$&$=$&$X_{Q^1}-3 $\\
$~\phantom{\Big|}X_{\overline{D^1}}$&$=$&$-X_{
H^\mathcal{D}}-X_{Q^1}+4+x $\\
$~\phantom{\Big|}X_{\overline{D^2}}$&$=$&$X_{
\overline{D^1}}-1 $\\
$~\phantom{\Big|}X_{\overline{D^3}}$&$=$&$
X_{\overline{D^1}}-1 $\\
$~\phantom{\Big|}X_{\overline{U^1}}$&$=$&$
X_{H^\mathcal{D}}-X_{Q^1}+9 $\\
$~\phantom{\Big|}X_{\overline{U^2}}$&$=$&$
X_{\overline{U^1}}-3 $\\
$~\phantom{\Big|}X_{\overline{U^3}}$&$=$&$
X_{\overline{U^1}}-5 $\\
$~\phantom{\Big|}X_{L^1}$&$=$&$\frac{1}{2}+
X_{H^\mathcal{D}}+\Delta^{\!H}$\\
$~\phantom{\Big|}X_{L^2}$&$=$&$X_{L^1} + 1-
\Delta^L_{31}+  3\zeta $\\
$~\phantom{\Big|}X_{L^3}$&$=$&$X_{L^1}\phantom{+1}~~
+\Delta^L_{31}$\\
$~\phantom{\Big|}X_{\overline{E^1}}$&$=$&$-X_{
H^\mathcal{D}}        {+~5} - X_{L^1} + x  $\\
$~\phantom{\Big|}X_{\overline{E^2}}$&$=$&$-X_{
H^\mathcal{D}}  {+~1} - X_{L^1} + x +
\Delta^L_{31} - 3\zeta      $ \\ 
$~\phantom{\Big|}X_{\overline{E^3}}$&$=$&$-
X_{H^\mathcal{D}}  \phantom{+~4} - X_{L^1} + 
x ~~-\Delta^L_{31}             $ \\ 
$~\phantom{\Big|}X_{\overline{N^{     1}}}     
$&$=$&$ \frac{1}{2}+\nu   $ \\ 
$~\phantom{\Big|}X_{\overline{N^{2}}}     $&$=$&$
 \frac{1}{2}+\nu+\Delta^{\overline{N}}_{21} $ \\
  & &\\ \hline
\end{tabular}
\caption{\label{Table2.5} 
  In addition to Table~\ref{Table2}, the constraints from the
  $\mu$-parameter $z=1$ and proton decay $y=0$ are imposed.
  Furthermore,  proton decay prefers $x=2,3$ over $x=0,1$.
  $\Delta^L_{31},~\zeta,~\Delta^{\!H},~\nu,~\Delta^{\overline{N}}_{21}$
  are integers.} 
\end{center}
\end{table}
%

%
%
%
\section{\label{nuPHe}Neutrino Phenomenology}
\cleqn
Our study of the neutrino sector is far more constrained than most
models in the literature.  This is because there is no GUT scale,
which is a factor of $\sim 100$ lower than $M_{grav}$, to suppress the
mass scale of the Majorana mass terms $\sim 10^{15}$ GeV, or
equivalently, boost the 
light neutrino masses to the required orders of magnitude.  In typical
seesaw models (see Refs.~\cite{ty,shelly,mgmrs,mohausen}), it is
achieved using an extra symmetry, such as gauged $U(1)_{B-L}$.
However, in our scenario there are no additional symmetries beyond 
the
$M\!S\!S\!M$ gauge groups and $U(1)_X$ nor additional mass scales beyond
$M_{grav}$ and $m_{3/2}$;  therefore 
the mass scales of right-handed neutrinos originate from 
$M_{grav}$, suppressed by powers of $\epsilon$. As our model
contains only two right-handed neutrinos, the mass of the lightest 
neutrino is zero.  The successful
neutrino phenomenology together with proton decay constraints
determine the $U(1)_X$ charge assignments down to 
four choices, see Tables~\ref{Table5}-\ref{Table8}.


Because this discussion is rather long, we have divided this section
into the following subsections.  In Section~\ref{subsec:mixing}, we
review our phenomenological understanding of neutrino mixings and
discuss their implications on $U(1)_X$ charge assignments.  
Phenomenology requires $\zeta = \Delta_{31}^L = -1$ as well as $z=1$,
thus justifying the GM mechanism for the $\mu$-parameter from a
completely different reasoning.  The resulting charge assignments are
shown in Table~\ref{Table3}.  Section~\ref{subsec:masses} is the
corresponding discussion of neutrino mass eigenvalues.  Here we
encounter different possibilities depending on whether 
$LL$, $LR$ and $RR$
entries of the neutrino mass matrices are 
induced by the GM mechanism,
schematically shown in Table~\ref{Table4}:  Section~\ref{921}
discusses cases \emph{1.)}, \emph{2.)}, and \emph{3.)}, while
Section~\ref{negative} discusses cases \emph{4.)}, \emph{5.)}, and
\emph{6.)}. We find successful
solutions to cases \emph{2.)} and \emph{6.)}.  The former case is 
similar to the standard seesaw scenario, and $U(1)_X$ charge assignments
are shown in 
Tables~\ref{Table5} and \ref{Table6}.  The latter case has the
right-handed neutrino masses from the GM mechanism and hence they are
present below the electroweak scale.   The $U(1)_X$ charge 
assignments are shown in Tables~\ref{Table7} and \ref{Table8}.

\subsection{\label{subsec:mixing}Neutrino Mixing}
If there are no filled up 
supersymmetric 
zeros  in  $\boldsymbol{G^{(U)}}$ and  
$\boldsymbol{G^{(D)}}$, then, see 
Ref.~\cite{lns1}, 
\beq\label{ccckkkmmm}
{V^{\!C\!K\!M}}_{ij}\sim\epsilon^{|X_{Q^i}-X_{Q^j}|}.
\eeq 
Analogously, if there are no filled up supersymmetric 
zeros in the mass matrices in the leptonic sector, one 
has
\beq\label{ckmfornu}
{U^{\!M\!N\!S}}_{ij}\sim\epsilon^{|X_{L^i}-X_{L^j}|},
\eeq
\emph{i.e.} a  symmetric   (with respect to the 
$\epsilon$-suppression) Maki-Nagakawa-Sakata (MNS) matrix 
(see Ref.~\cite{mns}).
Phenomenology 
suggests, 
see  \emph{e.g.} Ref.~\cite{bk} (using Refs.~\cite{satoya,vassini})
\beqn\label{mmnnss}
{\boldsymbol{U^{\!M\!N\!S}}}~\sim~\left(\begin{array}{lll}
1 & \eps & \eps \\
\eps & 1 & 1 \\
\eps & 1 & 1 \end{array}\right),
\end{eqnarray}
with possibly higher exponents of $\epsilon$ in the 
(1,3)-element. Comparing with Eq.~(\ref{ckmfornu}) gives
\beq
|X_{L^1}-X_{L^2}|=|X_{L^1}-X_{L^3}|=1,~~~
|X_{L^2}-X_{L^3}|=0,
\eeq
so that 
\beq\label{dd1}
\Delta^L_{21}=\Delta^L_{31}=\pm 1. 
\eeq
Combining this with 
Eq.~(\ref{ddzz}) gives 
\beq\label{zzeettaa}
\zeta=\frac{\pm2-z}{3}.
\eeq
Since $\zeta$ has to be integer, one is left with 
\beq\label{nsz}
\zeta=\Delta^L_{31}=-1~~~\mbox{and}~~~z=1.
\eeq
It is interesting to notice that the  MNS phenomenology 
combined with  the requirement that there are no filled up
 superymmetric zeros and guaranteeing $R_p$  the way 
advocated here \emph{predicts} $z=1$, \emph{i.e.} the necessity 
to have the $\mu$-term generated via GM! {We 
also looked at the 
more general case with the possibility of supersymmetric 
zeros in  $\boldsymbol{G^{(N)}}$ and  
$\boldsymbol{G^{(E)}}$, not leading
to a  substantially different result.
For completeness the calculations
generalizing the lower case of the lower
left-hand corner of Table~\ref{Table4} are 
given in  Appendix~\ref{mitsusy0}.} 
Plugging Eq.~(\ref{nsz}) and $y=0$ into Table~\ref{Table2}
gives Table~\ref{Table3}. The only non-neutrino parameters
left unfixed are $x$ and $\Delta^{\!H}$.

{}From Eq.~(\ref{nsz}) one can observe furthermore that there are no 
supersymmetric zeros for the superfield operators $QQQL$
and  $\overline{U}\overline{U}\overline{D}\overline{E}$ 
so that the canonicalization of the K\"ahler potential 
does not affect the order of $\eps$-suppression. 
With the results of Ref.~\cite{hr}, also translated to
 the mass matrices of charged leptons and 
light neutrinos we   find that the powers of $\epsilon$ 
for $QQQL$ and  $\overline{U}\overline{U}\overline{D}\overline{E}$
  are again not changed when going to the
mass basis.\footnote{This was assumed to be true in Ref.~\cite{ky},
  while we explicitly verified it.} 
With Eq.~(\ref{nsz}) and
\beqn
0.186^{10}\approx5\times10^{-8},  & & 0.186^{10}
\approx5\times10^{-8} ,\nonumber\\
0.198^{11}\approx 2\times10^{-8}, & & 0.198^{12}
\approx 4\times10^{-9} ,\nonumber\\
0.210^{12}\approx 7\times10^{-9},  & & 0.210^{14}
\approx 3\times10^{-10}, \nonumber\\
0.221^{13}\approx 3\times10^{-9},  & &0.221^{16}
\approx 3\times10^{-11} \eeqn
together with Eq.~(\ref{kackizacki}), one finds that the cases 
with $x=0,1$  are ruled out,
only $x=2,3$ are viable, while $x=3$ is allowed even for 
$m_{squark}=100$~GeV.\footnote{In the language of 
Ref.~\cite{dt}, the model with $\Delta^L_{31} = \zeta = -1$, $z=1$,
and $y=0$ is classified as 
\emph{(no)$_{h.o.}$}.}

Will we be able to have an $X$-charge assignment such that
 no hidden sector fields are needed in order to cancel the
 anomalies of $U(1)_X$ with itself and gravity? 
Table~\ref{Table1} 
and Eq.~(\ref{ggx}) give 
\beq
\mathcal{A}_{GGX}=59 - 3 X_{H^\mathcal{D}} + 3X_{L^1}  
+ 12 x + 8 z + \Delta^L_{21} + \Delta^L_{31}+\sum_I 
X_{\overline{N^I}}+\mathcal{A}_{GGX}^{hidden~sector},
\eeq
focusing on our  $R_p$-conserving scenario one obtains
\beq
\mathcal{A}_{GGX}=\frac{121}{2}+ 12 x + 9z + 
3\Delta^{\!H} +3\zeta+\sum_I X_{\overline{N^I}}+
\mathcal{A}_{GGX}^{hidden~sector}.
\eeq
With Eq.~(\ref{nsz})  and assuming no $X$-charged 
hidden fields we get
\beq\label{gvsc}
\mathcal{A}_{GGX}=66+\mbox{$\frac{1}{2}$}+12x+ 3
 \Delta^{\!H}+\sum_I X_{\overline{N^I}}.
\eeq
With Eq.~(\ref{cc}) and $x=3$ (and hence
$\mathcal{A}_{CCX}=30/2$),
one finds that
${360}/{k_C}$ has to be a half-odd-integer number, 
which is only given for $k_C=16,48,80,144,240,720$, 
resulting in $~3\Delta^{\!H}+\sum_I X_{\overline{N^I}}=
-80,-95,-98,-100,-101,-102$,~ respectively. 
Analogously for 
$x=2$ (and hence
$\mathcal{A}_{CCX}=27/2$), we obtain  $k_C=8,24,72,216,648$ and  
$~3\Delta^{\!H}+\sum_I X_{\overline{N^I}}=-
50,-77,-86,-89,-90$,~ respectively.
We consider this to be highly 
unlikely (confirmed in the next Subsections), since 
it requires extremely large 
$X$-charges. Moreover, such charge assignments would require the
neutrino Dirac mass matrix to be generated by the GM mechanism and
hence the neutrino masses come out too small (see the next subsection).
So to have only $M\!S\!S\!M$ superfields 
and  $\{A,\overline{N^I}\}$ to be 
$X$-charged is not possible.\footnote{Alas, 
this would have enabled us to determine 
$\mathcal{A}_{X\!X\!X}$ and thus $k_X$ and thus $g_X$.} 
The rest of the goals mentioned in Section~2 will be achieved 
however.
 
It should be mentioned that phenomenology might 
also suggest  the so called anarchical scenario, see 
Refs.~\cite{anarchy1,
hm,anarchy3}, \emph{i.e.} instead of Eq.~(\ref{mmnnss})
one has
\beqn
{\boldsymbol{U^{\!M\!N\!S}}}~\sim~\left(\begin{array}{lll}
1 & 1 & 1 \\
1 & 1 & 1 \\1 & 1 & 1 \end{array}\right).\end{eqnarray}
However, this is not compatible with  $z=1$ in combination 
with  Eq.~(\ref{ddzz}). 

\begin{table}
\begin{center}
\vspace{-1cm}
\begin{tabular}{|rcl|}
\hline
  & & \\
$~~~\phantom{\Big|}X_{H^\mathcal{D}}$
 &$=$&$   \frac{2}{5} - \frac{39 - 4\,x\,
      \left( 6 + x - {\Delta^{\!H}} \right)  + 
     30\,{\Delta^{\! H}}}{10\,\left( 7 + x \right) }  $\\
$~\phantom{\Big|}X_{H^\mathcal{U}}$
 &$=$&$-1-\phantom{\Big|}X_{H^\mathcal{D}}  $\\
$~\phantom{\Big|}X_{Q^1}$&$=$&$\frac{1}{3}\Big(
      \frac{23}{2} - X_{H^\mathcal{D}} + x  - 
\Delta^{\!H}\Big)$\\
$~\phantom{\Big|}X_{Q^2}$&$=$&$X_{Q^1}-1 $\\
$~\phantom{\Big|}X_{Q^3}$&$=$&$X_{Q^1}-3 $\\
$~\phantom{\Big|}X_{\overline{D^1}}$&$=$&$-X_{
H^\mathcal{D}}-X_{Q^1}+4+x $\\
$~\phantom{\Big|}X_{\overline{D^2}}$&$=$&$X_{
\overline{D^1}}-1 $\\
$~\phantom{\Big|}X_{\overline{D^3}}$&$=$&$
X_{\overline{D^1}}-1 $\\
$~\phantom{\Big|}X_{\overline{U^1}}$&$=$&$
X_{H^\mathcal{D}}-X_{Q^1}+9 $\\
$~\phantom{\Big|}X_{\overline{U^2}}$&$=$&$
X_{\overline{U^1}}-3 $\\
$~\phantom{\Big|}X_{\overline{U^3}}$&$=$&$
X_{\overline{U^1}}-5 $\\
$~\phantom{\Big|}X_{L^1}$&$=$&$\frac{1}{2}+
X_{H^\mathcal{D}}+\Delta^{\!H}$\\
$~\phantom{\Big|}X_{L^2}$&$=$&$X_{L^1} -1 $\\
$~\phantom{\Big|}X_{L^3}$&$=$&$X_{L^1}-1 $\\
$~\phantom{\Big|}X_{\overline{E^1}}$&$=$&$-X_{
H^\mathcal{D}}        {+~5} - X_{L^1} + x  $\\
$~\phantom{\Big|}X_{\overline{E^2}}$&$=$&$-X_{
H^\mathcal{D}}  {+~3} - X_{L^1} + x      $ \\ 
$~\phantom{\Big|}X_{\overline{E^3}}$&$=$&$-
X_{H^\mathcal{D}}  {+~1} - X_{L^1} + x   $ \\ 
$~\phantom{\Big|}X_{\overline{N^{     1}}}     
$&$=$&$ \frac{1}{2}+\nu   $ \\ 
$~\phantom{\Big|}X_{\overline{N^{2}}}     $&$=$&$
 \frac{1}{2}+\nu+\Delta^{\overline{N}}_{21} $ \\
  & &\\ \hline
\end{tabular}
\caption{\label{Table3} 
In addition to Table~\ref{Table2.5}, we 
required successful neutrino
  mixings, {\it \emph{i.e.}}\/, 
$\Delta^L_{31}=\zeta=-1$.  The remaining
  parameters $\Delta^{\!H},~\nu, ~\Delta^{\overline{N}}_{21} $ are
  integers, while $x=2$ or 3 to satisfy proton decay constraints.}
\end{center}
\end{table}
%

%
%
%
%
\subsection{\label{subsec:masses}Neutrino Masses}
After looking at the mixing, let us now investigate the 
relationship between  the neutrino mass spectrum and the 
$X$-charges. For future reference we here state the
experimental status, allowing for three possible neutrino mass
 solutions, see \emph{e.g.} Ref.~\cite{nirgg} and 
references therein:
\begin{itemize}
\item ``hierarchical''($m_{\nu^3}$ is much larger than 
$m_{\nu^2}$, which is much larger than $m_{\nu^1}$),
\beqn\label{hir}
{m_{\nu^2}}^2-{m_{\nu^1}}^2\phantom{-}&\sim&
\phantom{-}7\times10^{-5}~
\mbox{eV}^2,\nonumber\\
{m_{\nu^3}}^2-{m_{\nu^2}}^2\phantom{-}&\sim&
\phantom{-}3\times10^{-3}~
\mbox{eV}^2,
\eeqn
\item ``inverse hierarchical''($m_{\nu^2}$ is minutely
 larger than $m_{\nu^1}$, which is much larger 
than $m_{\nu^3}$; this is not possible in
our scenario),
\beqn\label{jnv}
{m_{\nu^2}}^2-{m_{\nu^1}}^2\phantom{-}&
\sim&\phantom{-}7\times10^{-5}~
\mbox{eV}^2,\nonumber\\
{m_{\nu^3}}^2-{m_{\nu^2}}^2\phantom{-}&
\sim&-3\times10^{-3}~
\mbox{eV}^2,\eeqn
\item ``quasi-degenerate'' (all $m_{\nu}$ are almost 
identical; this is not possible in our scenario).
\end{itemize}

The scenario sketched so far (with $x=2,3$, 
$y=0$, $z=1$, $\zeta=-1$, $\Delta^L_{31}=-1$)  generates 
a superpotential (neglecting the tiny contributions of 
the GM mechanism to $\boldsymbol{G^{(U,D,E)}}$) with
[for $\bsym{G^{(E)}}$ see Eq.~(\ref{59999})]
\begin{eqnarray}\label{superpotmssm}
\mcal{W}^{M\!S\!S\!M}~&=&~{g^{(U)}}_{\!ij}~   
\left(\begin{array}{lll}\eps^8 &\eps^{5} &\eps^{3} \
\\\eps^{7} &\eps^4 &\eps^2 \\\eps^5 &\eps^2 & 1
\end{array}\!\!\right)_{\!ij}~Q^i~{H^{\mcal{U}}}~
\overline{U^j}~+~m_{3/2}~~\widetilde{g^{(\mu)}}~~
\epsilon~{H^{\mcal{D}}}~
{H^{\mcal{U}}}\nonum\\
&&+~{g^{(D)}}_{\!ij}~\eps^{{2~\mbox{\scriptsize or}~3}}~
    \left(\begin{array}{lll}\eps^{4} &\eps^{3} &\eps^3 \\
\eps^{3} &\eps^2 &\eps^2\\\eps & 1 & 1\end{array}
\right)_{\!ij}~Q^i~
{H^{\mcal{D}}}~\overline{D^j}~ \nonumber \\
&&+~{g^{(E)}}_{\!ij}~
\eps^{2~\mbox{\scriptsize or}~3}~\left(\begin{array}{lll}\eps^{5} &
\eps^{3} &
\eps \\\eps^{4} &\eps^2 & 1 
\\\eps^4 &\eps^2 & 1\end{array}\right)_{\!ij}~L^i~
{H^{\mcal{D}}}~
\overline{E^j},
\end{eqnarray}
and
\begin{eqnarray}\label{superpotnu}
\mcal{W}^{(\nu)}&=&\frac{1}{2}~\Big(\Theta[X_{\overline{N^I}}+
X_{\overline{N^J}}]~~M_{grav}~~\gamma_{\!IJ}~~
\eps^{X_{\overline{N^I}}+X_{\overline{N^J}}}\nonum\\
&&\qquad+~m_{3/2}\cdot
\widetilde{\gamma}_{\!IJ}~~\eps^{|X_{\overline{N^I}}
+X_{\overline{N^J}}|}\Big)~~\overline{N^I}~~
\overline{N^J}~~\nonum\\
&&+~~\Big(\Theta[X_{L^i}+X_{H^\mathcal{U}}+
X_{\overline{N^J}}]~~{g^{(N)}}_{\!iJ}~~\eps^{X_{L^i}+
X_{H^\mathcal{U}}+X_{\overline{N^J}}}~~\nonum\\
&&\qquad +~~\frac{m_{3/2}}{M_{grav}}\cdot
{\widetilde{g^{(N)}}}_{\!iJ}~~\eps^{|X_{L^i}+
X_{H^\mathcal{U}}+X_{\overline{N^J}}|}\Big)~~{L^i}~~
{H^{\mcal{U}}}~~\overline{N^J} \nonum\\
&&+~~\frac{1}{2}~\Big(\Theta[X_{L^i}+X_{H^\mathcal{U}}+X_{L^j}
+X_{H^\mathcal{U}}]~~\frac{{\psi_{\!ij}}}    {M_{grav}}
~~\eps^{X_{L^i}+X_{H^\mathcal{U}}+X_{L^j}+
X_{H^\mathcal{U}}                }~~\nonum\\
&&\qquad+~~\frac{m_{3/2}}{M_{grav}}\cdot
\frac{{\widetilde{\psi}}_{\!ij}}{M_{grav}}~~
\eps^{|X_{L^i}+X_{H^\mathcal{U}}+X_{L^j}+
X_{H^\mathcal{U}}|}\Big)~~{L^i}~~{H^{\mcal{U}}}~~{L^j}
~~{H^{\mcal{U}}};~~~~~~~
\end{eqnarray}
summation over repeated indices is implied.
The two equations above do not contain any factors of 
$\Omega[...]$, because by construction all $R_p$-conserving 
terms  have integer $X$-charge. 
We will in turn investigate the different possibilities
 for generating the mass terms, as given in 
Table~\ref{Table4}.
  With $y=0$ 
all exponents in $\boldsymbol{G^{(U,D,E)}}$ are positive.
 From this one may feel inspired to assume \emph{either}
 that all exponents in the mass terms of the neutrinos are
 positive (the case in the lower right-hand corner of 
Table~\ref{Table4}),
 \emph{or} (less restrictive) simply that for a given 
array of neutrino coupling constants all exponents are 
either negative or positive.

\begin{table}$~$\\
\renewcommand{\arraystretch}{1.3}
  \centering
  \begin{tabular}{|c|lc|lc|}
    \hline
    & \multicolumn{2}{c|}{$X_{L^i}+X_{H^\mathcal{U}}<0$}
    & \multicolumn{2}{c|}{$X_{L^i}+X_{H^\mathcal{U}}>0$}\\ \hline
    & & & \multicolumn{2}{l|}{\emph{5.)} $X_{L^i}+X_{H^\mathcal{U}} <
      -X_{\overline{N^I}}$} \\ 
    & & & $\boldsymbol{M_{LL}^{Maj}}$: & FN\\
    & \emph{4.)} & & $\boldsymbol{M_{LR}^{Dirac}}$: & GM\\
    $X_{\overline{N^I}}$ & $\boldsymbol{M_{LL}^{Maj}}$:& GM 
    & $\boldsymbol{M_{RR}^{Maj}}$:&GM\\     \cline{4-5}
    $<0$ & $\boldsymbol{M_{LR}^{Dirac}}$: & GM & 
    \multicolumn{2}{l|}{\emph{6.)} $X_{L^i}+X_{H^\mathcal{U}} \geq
      -X_{\overline{N^I}}$}\\
    & $\boldsymbol{M_{RR}^{Maj}}$: & GM & $\boldsymbol{M_{LL}^{Maj}}$: & FN\\
    & & & $\boldsymbol{M_{LR}^{Dirac}}$: & FN\\
    & & & $\boldsymbol{M_{RR}^{Maj}}$: & GM \\ \hline
    & \multicolumn{2}{l|}{\emph{3.)} $X_{\overline{N^I}} < 
      -X_{L^i}-X_{H^\mathcal{U}}$} & & \\ 
    & $\boldsymbol{M_{LL}^{Maj}}$: & GM & &\\
    & $\boldsymbol{M_{LR}^{Dirac}}$: & GM & \emph{1.)} & \\
    $X_{\overline{N^I}}$ & $\boldsymbol{M_{RR}^{Maj}}$: & FN 
    & $\boldsymbol{M_{LL}^{Maj}}$:& FN 
    \\     \cline{2-3}
    $>0$  & \multicolumn{2}{l|}{\emph{2.)} $X_{\overline{N^I}} \geq 
      -X_{L^i}-X_{H^\mathcal{U}}$} & $\boldsymbol{M_{LR}^{Dirac}}$: & FN\\
    & $\boldsymbol{M_{LL}^{Maj}}$: & GM & $\boldsymbol{M_{RR}^{Maj}}$: & FN\\
    & $\boldsymbol{M_{LR}^{Dirac}}$: & FN & & \\
    & $\boldsymbol{M_{RR}^{Maj}}$: & FN & &\\ \hline
  \end{tabular}
\renewcommand{\arraystretch}{1}
\caption{\label{Table4} The Majorana mass  of the 
left-handed neutrinos 
is denoted by $\boldsymbol{M_{LL}^{Maj}}$, the one for the
 right-handed neutrinos is given by $\boldsymbol{
M_{RR}^{Maj}}$, and the Dirac mass  is  $\boldsymbol{
M_{LR}^{Dirac}}.$ We work with two right-handed neutrinos.}
\end{table}

Are the cases sketched in Table~\ref{Table4} 
 Majorana or pseudo
 Dirac neutrinos? One has pseudo Dirac neutrinos if 
$\boldsymbol{M_{LR}^{Dirac}}\gg\boldsymbol{M_{LL}^{Maj}},
\boldsymbol{M_{RR}^{Maj}}$. We will investigate 
case by case
 (starting in the lower right-hand corner, proceeding 
anticlockwise) whether this condition can be met when  
$\boldsymbol{M_{LR}^{Dirac}}$ is generated via the FN
mechanism, since   
$\boldsymbol{M_{LR}^{Dirac}}$  generated by the GM 
mechanism produces 
neutrino masses which are  too small, one has
that
$[{M_{LR}^{Dirac}}]_{ij}\approx
\epsilon^{|X_{L^i}+X_{\overline{N^j}}+X_{H^{\cal 
      U}}|} \, \langle 
H^\mathcal{U}\rangle ~m_{3/2}/M_{grav}\leq 10^{-5}$ eV.

%
%
%
%
\subsubsection{\label{921}Positive $\boldsymbol{
X_{\overline{N^I}}}$}
First we will take the $X$-charges of all right-handed 
neutrino superfields to be positive.  This ensures that
 the scalar component of $A$ acquires a VEV, since its  
$X$-charge is 
negative 
and $\xi_X$ is positive. This way   the 
$D_{\!X}$-term does not acquire a VEV and supersymmetry is
 not broken by the DSWW mechanism, see 
Ref.~\cite{dsw0,dsw,ads,ads2}, at a much too high 
energy  scale. 

Of course, guaranteeing a VEV for $A$ 
does not automatically guarantee that the scalar 
components
 of the right-handed neutrino superfields do not get 
VEVs -- this we  simply have to postulate, in order  to 
conserve $R_p$ and to have only one flavon field. \\

\emph{1.)} We now consider the case in the lower right 
corner of Table~\ref{Table4}. All $\Theta[...]$ can 
be dropped:
\begin{eqnarray}
\mcal{W}^{(\nu)}&=&M_{grav}~~\frac{\gamma_{\!IJ}}{2}~~\eps^{
X_{\overline{N^I}}+X_{\overline{N^J}}}~~\overline{N^I}
~~\overline{N^J}~~+~~{g^{(N)}}_{\!iJ}~~\eps^{X_{L^i}+
X_{H^\mathcal{U}}+X_{\overline{N^J}}}~~{L^i}~~{H^{
\mcal{U}}}~~\overline{N^J} \nonum\\
&~&+~~\frac{{\psi_{\!ij}}}{2~M_{grav}}~~\eps^{X_{L^i}+
X_{H^\mathcal{U}}+X_{L^j}+X_{H^\mathcal{U}}}~~{L^i}~~{
H^{\mcal{U}}}~~{L^j}~~{H^{\mcal{U}}},~~~~~~~
\end{eqnarray}
for short (dropping all generational indices) 
\begin{eqnarray}\label{nochnenlabel}
\mcal{W}^{(\nu)}&=&M_{grav}~~\overline{\boldsymbol{N}}^T
~~\frac{\boldsymbol{\Gamma}}{2}~~\overline{\boldsymbol{N}}~~+
~~\boldsymbol{L}^T~~\boldsymbol{G^{(N)}}~~{H^{\mcal{U}}}
~~\overline{\boldsymbol{N}} ~~\nonum\\&~&~~~~~~+~~
\frac{1}{M_{grav}}~~\boldsymbol{L}^T~~\frac{\boldsymbol{\Psi}}{2}
~~\boldsymbol{L}~~{H^{\mcal{U}}}~~{H^{\mcal{U}}}.
\eeqn
We  get a Lagrangian with ($\boldsymbol{n_{L/R}}$ 
are left/right-handed neutrinos in the interaction basis)
\beq
\mathcal{L}\supset \langle H^{\mathcal{U}}\rangle~
\bsym{n_L}^T~\bsym{G^{(N)}}~\overline{\bsym{n_R}}~+~
M_{grav}~\overline{\bsym{n_R}}^T~\frac{\bsym{\Gamma}}{2}~
\overline{\bsym{n_R}}~+~\frac{\langle H^{\mathcal{U}}
\rangle^2}{M_{grav}}~\bsym{n_L}^T~\frac{\bsym{\Psi}}{2}~
\bsym{n_L},
\eeq
which equals
\beqn
\mathcal{L}&\supset&\frac{1}{2}~\left( \bsym{n_L}^T, 
~~\overline{\bsym{n_R}}^T \right)\cdot\left(
\begin{array}{ccc}{\langle H^{\mathcal{U}}\rangle^2}~
\frac{\bsym{\Psi}}{M_{grav}} &~& {\langle 
H^{\mathcal{U}}\rangle}~\bsym{G^{(N)}}\\ ~&~&~\\ 
\langle H^{\mathcal{U}}\rangle~\bsym{G^{(N)}}^T &~&
M_{grav}~~\bsym{\Gamma}\end{array}\right)\cdot
\left(\begin{array}{c}\bsym{n_L} \\ ~\\ 
\overline{\bsym{n_R}} \end{array} \right).~~~~~~~~
\label{notbigbookbutbigequation}
\eeqn
For the Majorana case, the masses of the light neutrinos
 are given by the positive square roots of the eigenvalues
 of $~~\big(\langle H^\mathcal{U}\rangle^2/
M_{grav}\big)^2~\boldsymbol{\mathcal{G}^{(\nu)}}
\boldsymbol{\mathcal{G}^{(\nu)}}^\dagger$,
with
\beq
\boldsymbol{\mathcal{G}^{(\nu)}}\equiv\boldsymbol{\Psi}
-\bsym{G^{(N)}}~\boldsymbol{\Gamma}^{-1}~\bsym{G^{(N)}}^T
\eeq
(for a derivation see Appendix~\ref{seeesooo}). Note 
that this 
statement is not the same as saying ``the masses of 
the light neutrinos are given by the absolute values  
of the eigenvalues of $\langle H^\mathcal{U}\rangle^2/
M_{grav}~\boldsymbol{\mathcal{G}^{(\nu)}}$''.  However,
  since the entries of $\boldsymbol{\mathcal{G}^{(\nu)}}$ 
 are hierarchically $\epsilon$-suppressed it is a good
 approximation, for a demonstration see Ref.~\cite{dt}; we
 shall assume here that this approximation has been made.
 One has\footnote{Proof:
\beqn
\lefteqn{
[{\bsym{\Gamma}}^{-1}~~{\bsym{\Gamma}}]_{IJ}=\sum_{K}[
\bsym{{\gamma}}^{-1}]_{_{IK}}~~{\gamma}_{_{KJ}}~~~\eps^{
-X_{\overline{N^I}} -X_{\overline{N^K}}}~~\eps^{X_{
\overline{N^K}} +X_{\overline{N^J}}}
}\nonumber\\
&&=\eps^{X_{\overline{N^J}} -X_{\overline{N^I}}}~~
\sum_{K}[\bsym{{\gamma}}^{-1}]_{_{IK}}~~{\gamma}_{_{KJ}}
~~=~~\eps^{X_{\overline{N^J}} -X_{\overline{N^I}}}~~
\delta_{IJ}
~~=~~\delta_{IJ}. \nonumber
\eeqn}
\beqn\label{hfhfhf}
[{{\bsym{\Gamma}}}^{-1}]_{IJ}~=~[\bsym{{\gamma}}^{-1}]_{
IJ}~~\eps^{-X_{\overline{N^I}} -X_{\overline{N^J}}}.
\eeqn
It follows that\footnote{Note that the $U(1)_X$ charges of 
the right-handed neutrinos practically drop out
  from the light neutrino masses and mixings.  This fact is
  well known, see \emph{e.g.} Ref.~\cite{satoya}.} 
\beqn
{\mathcal{G}^{(\nu)}}_{ij}~=~\epsilon^{X_{L^i}+X_{L^j}+
2X_{H^\mathcal{U}}}~~\Big(\psi_{ij}~-~\sum_{K,L}~~ 
{g^{(N)}}_{iK}~~[\bsym{{\gamma}}^{-1}]_{KL}~~{
g^{(N)}}_{jL}\Big).
\eeqn
{Note that this result also holds when 
there are supersymmetric zeros (as long as 
$\bsym{\Gamma}$ 
is invertible), one just has to replace 
${\gamma}_{_{IJ}}$ by  ${\gamma}_{_{IJ}}\times
\Theta[X_{\overline{N^I}}+X_{\overline{N^J}}]$, and 
likewise for ${g^{(N)}}_{iJ}$.} 
Since we are considering the case where 
$X_{L^i}+X_{H^\mathcal{U}}>0$, we have
$\eps^{X_{L^i}+X_{L^j}+2X_{H^\mathcal{U}}}<1$. 
Thus
\beq
m_{\nu}<\frac{\langle H^\mathcal{U}\rangle^2  }{M_{grav}}
\approx 1\times10^{-5}~\mbox{eV},
\eeq
which is smaller than the experimentally
required  $~\Delta{m_{\nu}}^2$, see 
Eqs.~(\ref{hir},\ref{jnv}). We conclude that light
 Majorana neutrinos 
from the case in the lower right-hand corner of 
Table~\ref{Table4} are 
ruled out by phenomenology.\\ 

Pseudo Dirac neutrinos are also not possible. One needs 
to fulfill two conditions which contradict each other:
\beqn
\bigg(\frac{\langle H^\mathcal{U} \rangle}{M_{grav}}
\bigg)^2~\epsilon^{X_{L^i}+X_{L^j}+2X_{H^\mathcal{U}}}
&\ll &\frac{\langle H^\mathcal{U} \rangle}{M_{grav}}~
\epsilon^{X_{L^i}+X_{H^\mathcal{U}}+X_{\overline{N^J}}},
\nonumber\\  & &\nonumber\\
\epsilon^{ X_{  \overline{N^I}}+ X_{\overline{N^J}} }&
\ll &\frac{\langle H^\mathcal{U} \rangle}{M_{grav}}~
\epsilon^{X_{L^i}+X_{H^\mathcal{U}}+X_{\overline{N^J}}}.
\eeqn
$~$\\

\emph{2.)} Now we consider the lower case in the lower 
left-hand corner  in 
Table~\ref{Table4}, first the Majorana case. $
\boldsymbol{M_{LL}^{Maj}}$ is suppressed by a factor 
of $m_{3/2}/M_{grav}$, compared to the one in \emph{1.)},
 so that we can neglect it. We arrive at 
\beq\label{grouchy}
{\mathcal{G}^{(\nu)}}_{ij}~=~-~\epsilon^{X_{L^i}+X_{L^j}+
2X_{H^\mathcal{U}}}~~\sum_{K,L}~~ {g^{(N)}}_{iK}~~
[\bsym{{\gamma}}^{-1}]_{KL}~~{g^{(N)}}_{jL}.
\eeq
Unlike the previous case, this time $X_{L^i}+
X_{H^{\mathcal{U}}}<0$, so that  $\eps^{{X_{L^i}+
X_{L^j}+2X_{H^\mathcal{U}}}  }>1$, and thus $m_\nu$ 
can be ``$\epsilon$-enhanced'' to agree with 
phenomenology. 

{}From the equation above it follows that
\beqn
\det\big[\bsym{\mathcal{G}^{(\nu)}}\big]=-\det\big[
\bsym{{g}^{(N)}}\bsym{{\gamma}}^{-1}\bsym{{g}^{(N)}}^T
\big]\cdot\eps^{6X_{H^\mathcal{U}}+2\sum_{i}X_{L^i}}.
\eeqn   
Since $\bsym{\gamma}$ is a $2\times2$ matrix and 
$\bsym{g^{(N)}}$ is a  $3\times2$ matrix,  the 
determinant of  $\bsym{{g}^{(N)}}\bsym{{\gamma}}^{-1}
\bsym{{g}^{(N)}}^T$   is zero, regardless of which 
values one has for the entries of $\boldsymbol{\gamma}$,
 $\boldsymbol{g^{(N)}}$
(this constrained seesaw mechanism was first proposed in
  Ref.~\cite{fgy}; for a model embedding  into a 
family symmetry  see Ref.~\cite{srabi}). Thus one of the 
three eigenvalues of  $\bsym{\mathcal{G}^{(\nu)}}$ is 
definitely zero, so that the lightest neutrino is 
massless;
 we can use   $~\Delta{m_{\nu}}^2~$ to determine the 
absolute masses of the two other neutrinos. Using
Eq.~(\ref{hir}) and\footnote{We easily obey  the limit on 
neutrino masses  from  WMAP, see Refs.~\cite{wmap1,wmap2},
 namely 
$$\sum m_\nu\leq 0.71~\mbox{eV}.$$}  
with $\lambda$ denoting the eigenvalues of the matrix in  
Eq.~(\ref{grouchy}),   for the hierarchical case one has
\beqn
\sqrt{7\times10^{-5}~}~\mbox{eV}~\sim~\frac{\langle 
H^\mathcal{U}\rangle^2}{M_{grav}}\times
\lambda_{middle},\nonumber\\
\sqrt{3\times10^{-3}~}~\mbox{eV}~\sim~\frac{\langle 
H^\mathcal{U}\rangle^2}{M_{grav}}\times\lambda_{max}.
\eeqn\label{848}
Hence with $x=3$, $\tan\beta$ is smallish (say, 3) and thus 
with $\langle H^\mathcal{U}\rangle\approx 164$~GeV, one finds
\beqn
\lambda_{middle}\sim746\approx0.221^{-4.4},
& &\lambda_{max}\sim4887\approx0.221^{-5.6},
\eeqn
\emph{i.e.} (see Appendix~\ref{kalk}) 
\beq
2(X_{L^2}+X_{H^\mathcal{U}})\approx-4.4, \qquad
2(X_{L^3}+X_{H^\mathcal{U}})\approx-5.6.
\eeq
For $x=2$ the value of  $\tan\beta$ is larger so that 
$\langle H^\mathcal{U}
\rangle$ is a bit closer to 174~GeV, and $\epsilon=0.210$,
 but the results
for the $X_{L^i}+X_{H^\mathcal{U}}$  are very similar.
We allow for the possibility of a mild fine-tuning 
such that \emph{e.g.} 
\beq
\det\underbrace{\left(\begin{array}{cc}\mathcal{O}(1)&
\mathcal{O}(1)\\\mathcal{O}(1)&\mathcal{O}(1)\end{array}
\right)}_{\equiv \boldsymbol{\mathcal{M}}}=\mathcal{O}(
\epsilon)~~~\mbox{or}~~~=\mathcal{O}(1/\epsilon),~~~
\mbox{rather than}~~~ =\mathcal{O}(1),
\eeq
so that the eigenvalues of $\mathcal{M}$ are $\big\{
\mathcal{O}(1)$, $\mathcal{O}(\epsilon)\big\}$ or 
$\big\{\mathcal{O}(1)$, $\mathcal{O}(1/\epsilon)\big\}$ 
rather than  $\big\{\mathcal{O}(1)$, $\mathcal{O}(1)
\big\}$.\footnote{That the $\mathcal{O}(1)$ coefficients
 cannot be completely random is generic 
to all models with an MNS matrix as given in Eq.~(\ref{mmnnss}), see  \emph{e.g.} Ref.~\cite{bk}. Note that a pair of eigenvalues 
$\big\{\mathcal{O}(1)$, $\mathcal{O}(\epsilon)\big\}$ 
can be due to either an accidental cancellation among   $\mathcal{O}(1)$ coefficients or small $\mathcal{O}(1)$ coefficients, while $\big\{\mathcal{O}(1)$, $\mathcal{O}(1/\epsilon)\big\}$  can only be due to large $\mathcal{O}(1)$ coefficients. }  So we  can approximate
\beq
2(X_{L^2}+X_{H^\mathcal{U}})\approx -4,-5,~~~
2(X_{L^3}+X_{H^\mathcal{U}})\approx -5,-6.
\eeq
Since the superfield operator $~L^iH^\mathcal{U}~$ 
violates $R_p$,
  $X_{L^i}+X_{H^\mathcal{U}}$ has to be half-odd-integer, 
 so
\beq
2(X_{L^2}+X_{H^\mathcal{U}})= 2(X_{L^3}+X_{H^\mathcal{U}}) =  -5
\eeq
(note also that Eq.~(\ref{dd1}) requires $X_{L^2}=X_{L^3}$).
This gives (the same textures were anticipated by
Refs.~\cite{jj,bgw,lr,bk})  
\beqn\label{mathcalgnu}
{\boldsymbol{\mathcal{G}^{(\nu)}}}~\sim~\eps^{-5}~
\left(\begin{array}{lll}
\eps^2 & \eps & \eps \\
\eps & 1 & 1 \\
\eps & 1 & 1 \end{array}\right)
\end{eqnarray}
and 
\beq
\Delta^{\!H}=-1.
\eeq
Using this in Table~\ref{Table3} one finds the 
complete set of
$X$-charges for $x=2,3$, displayed in  
Tables~\ref{Table5},\ref{Table6}.\footnote{Actually 
one of the 
$X_{\overline{N^I}}$ 
may be $3/2$, as shown in Appendix~\ref{mitsusy0}.}
\begin{table}
\begin{center}
\begin{tabular}{|c|}
\hline $\phantom{\Bigg|}X_{\overline{N^1}}~\geq~
\frac{5}{2},~~~~$  $X_{\overline{N^2}}~\geq~\frac{5}{2}$
  \\
\hline
\end{tabular}~
\end{center}
\begin{center}
\begin{tabular}{|c|}
\hline $\phantom{\Bigg|}X_{H^\mathcal{D}}~=~
\frac{11}{10},~~~~$  $X_{H^\mathcal{U}}=~-~
\frac{21}{10}$  \\
\hline
\end{tabular}~
\end{center}
\begin{center}
\begin{tabular}{|c|c|c|c|c|c|}
\hline
\bf{Generation}\phantom{$\Big|$} $\bsym{i}$~ &~~~ $
\bsym{X_{\!Q^i}}$~~~ &~~~ $\bsym{X_{\!\overline{D^i}}}$
~~~ &~~~ $
\bsym{X_{\!\overline{U^i}}}$~~~ &~~~ $\bsym{X_{\!L^i}}$
~~~ &~~~ $
\bsym{X_{\!\overline{E^i}}}$~~~       \\\hline
1\phantom{\Bigg|} & $\frac{67}{15}$ &$\frac{13}{30}$ &
$\frac{169}{30}$ &$\frac{3}{5} $ &$\frac{53}{10} $
  \\  \hline
2\phantom{\Bigg|} & $\frac{52}{15}$ &$-\frac{17}{30}
\phantom{-}$ &$\frac{79}{30} $ &$-\frac{2}{5}\phantom{-}
 $ &$
\frac{33}{10} $  \\  \hline 
3\phantom{\Bigg|} & $\frac{22}{15}$ &$-\frac{17}{30}
\phantom{-}$ &$\frac{19}{30}$ &$-\frac{2}{5}\phantom{-}
 $ &$
\frac{ 13}{10}  $ \\  \hline 
\end{tabular}
\end{center}
\caption{Solution to case \emph{2.)} ($\Delta^{\!H}=-1$) with $x=2$.}
\label{Table5}
\end{table}
\begin{table}
\begin{center}
\begin{tabular}{|c|}
\hline $\phantom{\Bigg|}X_{\overline{N^1}}~\geq~
\frac{5}{2},~~~~$  $X_{\overline{N^2}}~\geq~\frac{5}{2}$
  \\
\hline
\end{tabular}~
\end{center}
\begin{center}
\begin{tabular}{|c|}
\hline $\phantom{\Bigg|}X_{H^\mathcal{D}}~=~
\frac{151}{100},~~~~$  $X_{H^\mathcal{U}}=~-~
\frac{251}{100}$  \\
\hline
\end{tabular}~
\end{center}
\begin{center}
\begin{tabular}{|c|c|c|c|c|c|}
\hline
\bf{Generation}\phantom{$\Big|$} $\bsym{i}$~ &~~~ $
\bsym{X_{\!Q^i}}$~~~ &~~~ $\bsym{X_{\!\overline{D^i}}}$
~~~ &~~~ $
\bsym{X_{\!\overline{U^i}}}$~~~ &~~~ $\bsym{X_{\!L^i}}$
~~~ &~~~ $
\bsym{X_{\!\overline{E^i}}}$~~~       \\\hline
1\phantom{\Bigg|} & $\frac{1399}{300}$ &$\frac{62}{75}$ &
$\frac{877}{150}$ &$\frac{101}{100} $ &$\frac{137}{25} $
  \\  \hline
2\phantom{\Bigg|} & $\frac{1099}{300}$ &$-\frac{13}{75}
\phantom{-}$ &$\frac{427}{150}$ &$\frac{1}{100} $ &$
\frac{ 87}{25} $  \\  \hline 
3\phantom{\Bigg|} & $\frac{499}{300}$ &$-\frac{13}{75}
\phantom{-}$ &$\frac{127}{150}$ &$\frac{ 1}{100} $ &$
\frac{ 37}{25}  $ \\  \hline 
\end{tabular}
\end{center}
\caption{Solution to case \emph{2.)} ($\Delta^{\!H}=-1$) with $x=3$.}
\label{Table6}
\end{table}
The maximum absolute value of the non-neutrino 
$X$-charges
  in Table~\ref{Table5} is $X_{\overline{U^1}}=5.63$,
 the minimum absolute value is $X_{{L^{2,3}}}=0.4$. 
For  Table~\ref{Table6}  one has $X_{\overline{U^1}}=
5.84$ and $X_{{L^{2,3}}}=0.01$; so among the $X$-charges
 are ratios of up to 500, but their values are below 10.
 \\

That we started with $\Delta^L_{21}=\Delta^L_{31}=-1$ 
was of course 
an inspired \emph{guess}  based on comparing 
Eq.~(\ref{mmnnss}) with the  hand waving 
Eq.~(\ref{ckmfornu}). So we have to check that the 
$X$-charges given above indeed lead to the 
  MNS matrix we used as a starting point.
This would justify our guess in hindsight.\footnote{An 
example that the starting  rule-of-the-thumb guess [to
apply Eq.~(\ref{ckmfornu}) to Eq.~(\ref{mmnnss})]
  does not automatically lead to the correct 
${\bsym{U^{\!M\!N\!S}}}$ in the end:  One might be  
willing to allow for a choice of the 
$\mathcal{O}(1)$-coefficients to be such that
$z=2$ is a possible. With Eqs.~(\ref{dd1},\ref{zzeettaa})
this gives $\Delta^L_{31}=1$, $\zeta=0$. Using 
 Eq.~(\ref{59999}) we get a  $\bsym{G^{(E)}}$
in which the (1,3)-entry dominates,
producing a ${\bsym{U^{\!M\!N\!S}}}$ which
is not in accord with Eq.~(\ref{mmnnss}).}
We get from Eq.~(\ref{superpotmssm}) and 
Eq.~(\ref{mathcalgnu}) that
\beqn
{\bsym{G^{(E)}}}{\bsym{G^{(E)}}}^\dagger~ \sim~ \eps^{2x}~
\left(\begin{array}{lll}
\eps^2 & \eps & \eps \\
\eps & 1 & 1 \\
\eps & 1 & 1 \end{array}\right),~~~~
{\boldsymbol{\mathcal{G}^{(\nu)}}}
{\boldsymbol{\mathcal{G}^{(\nu)}}}^\dagger~ \sim ~
\eps^{-10}~
\left(\begin{array}{lll}
\eps^2 & \eps & \eps \\
\eps & 1 & 1 \\
\eps & 1 & 1 \end{array}\right),\label{gg2}
\end{eqnarray}
so the two matrices which make up   
${\bsym{U^{\!M\!N\!S}}}$
both have a structure as in   Eq.~(\ref{mmnnss}).
To \emph{schematically} see this, consider the mass matrix 
in Eq.~(\ref{h-zwei}), dropping $\bsym{\Psi}$,  
$\bsym{\Gamma}$, $\bsym{G^{(N)}}$.
It is diagonalized by the matrix given in 
Eq.~(\ref{sowasvonkeinenbockmehr}), with its off-diagonal
 blocks approximated in Eq.~(\ref{h-drei}). Now replace 
all $\eta$ by $\epsilon$, one finds that
\beq
\left(\begin{array}{ll} \eps^2 & \eps\\
\eps & 1  \end{array}\right)~~\mbox{is diagonalized by}~~
\left(\begin{array}{ll} 1 & \eps\\
\eps & 1  \end{array}\right),
\eeq   
to be compared with Eq.~(\ref{gg2}) and  Eq.~(\ref{mmnnss}).

{}From the Tables~\ref{Table5},\ref{Table6}  we  
furthermore get that $3\Delta^{\!H}+\sum_IX_{\overline{N^I}}$
 does not allow for the Green-Schwarz anomaly cancellation
of $\mcal{A}_{GGX}$,
 [\emph{c.f.}  Eq.~(\ref{gvsc}) and the text below it]. So we are 
forced to require the existence of at least one 
$X$-charged superfield in the hidden   sector.\\

Pseudo Dirac neutrinos are possible, but require very 
large
 $X_{\overline{N^I}}$. As a toy model, consider
 the one-generational case. One of the two conditions 
not to  have Majorana masses is
\beq
\langle H^\mathcal{U} \rangle~ 
\eps^{X_{L}+X_{H^\mathcal{U}}+
X_{\overline{N}}}~\gg~ M_{grav}~ \eps^{2X_{\overline{N}}},
\eeq 
so that we need
\beq\label{hefgbvsb}
24 + X_{L}+X_{H^\mathcal{U}}~
<~{X_{\overline{N}}}.
\eeq
Phenomenology requires
\beq
\langle H^\mathcal{U} \rangle~ 
\eps^{X_{L}+X_{H^\mathcal{U}}+
X_{\overline{N}}}\sim10^{-2}~\mbox{eV},
\eeq 
hence
\beq\label{twenty-one}
X_{L}+X_{H^\mathcal{U}}+
X_{\overline{N}}\sim 21.
\eeq 
Using this in Eq.~(\ref{hefgbvsb}) gives  
$X_{\overline{N}}>\frac{45}{2}$. We will not go 
into more  detail, except for completeness stating the 
formul\ae\ with
which to relate the  Dirac  masses with the 
$X$-charges. The Dirac masses are
  $\langle H^\mathcal{U}\rangle$ times the positive 
square roots of the eigenvalues of
$\bsym{G^{(N)}}\bsym{G^{(N)}}^\dagger$. Since 
$\bsym{G^{(N)}}$
is a $3\times2$ matrix, the determinant of 
$\bsym{G^{(N)}}\bsym{G^{(N)}}^\dagger$ is zero, 
so one of 
the three eigenvalues is zero (so again one of the 
neutrinos
 is massless).  The other two eigenvalues
are equal to the two eigenvalues of the  non-singular 
$2\times2$ matrix
$\bsym{G^{(N)}}^\dagger\bsym{G^{(N)}}$. The powers of
$\epsilon$ of their square roots are  given as
\beqn
\mbox{min}\{X_{\overline{N^1}},X_{\overline{N^2}}\}~+~
X_{H^\mathcal{U}}~+~\mbox{min}\{X_{L^1},X_{L^2},X_{L^3}\},
\nonumber\\
\mbox{max}\{X_{\overline{N^1}},X_{\overline{N^2}}\}~+~
X_{H^\mathcal{U}}~+~\mbox{middle}\{X_{L^1},X_{L^2},
X_{L^3}\}.
\eeqn
$~$\\

\emph{3.)} Now we discuss the upper case in the lower 
left-hand corner  in 
Table~\ref{Table4}.  The Majorana case gives that  
the mass matrix 
of the light neutrinos is to lowest order $\propto\frac{\langle H^\mathcal{U}\rangle^2~m_{3/2}}{{M_{grav}}^2}$,
which is far too small. As explained earlier, the  
pseudo Dirac 
masses are not phenomenologically viable in this case, 
either. 

%
%
\subsubsection{\label{negative}Negative $\boldsymbol{X_{\overline{N^I}}}$}
Now we consider the case with  ${
X_{\overline{N^I}}}<0$, which is less appealing than 
the previous
 one, because a VEV of $A$ is no longer guaranteed.\\

\emph{4.)} First the upper left-hand corner in 
Table~\ref{Table4}.
 The 
Majorana case is similar to the one presented  in  
\emph{1.)}, but suppressed by an additional factor of
 $m_{3/2}/M_{grav}$, and thus the masses of the light 
neutrinos are far too small. As explained earlier,  
pseudo Dirac masses are not phenomenologically viable 
in this case, either.      \\

\emph{5.)} Now the upper case in the upper right-hand
 corner in Table~\ref{Table4}. The Majorana case 
gives that the 
mass matrix
 of the light neutrinos is
to lowest order 
\beq
{\psi}_{ij}~~\eps^{X_{L^i}+2X_{H^\mathcal{U}}+X_{L^j}}~~
\frac{\langle H^\mathcal{U}\rangle^2}{{M_{grav}}},
\eeq
which is too small. As explained earlier,  pseudo Dirac 
masses 
are not phenomenologically viable in this case, either.\\ 

\emph{6.)} Now the lower case in the upper right corner 
in Table~\ref{Table4}. We can neglect 
$\boldsymbol{M_{LL}^{Maj}}$ just
as in case \emph{1.)}.  We suppose that  the Majorana case 
does make sense, \emph{i.e.} we need that
$~~\langle H^\mathcal{U} \rangle\cdot {g^{(N)}}_{iJ}\cdot
\eps^{X_{L^i}+X_{H^\mathcal{U}}+{X_{\overline{N^J}}}}$ 
is much smaller than $m_{3/2}\cdot{\widetilde{\gamma}}_{IJ}
\cdot\eps^{|X_{\overline{N^I}}+X_{\overline{N^J}}|}$, in 
other words we  work with 
\beq\label{blc}
0<-3X_{\overline{N^I}}<
 X_{L^i}+X_{H^\mathcal{U}}.
\eeq 
Keeping in mind  that for negative $X_{\overline{N^I}}$ one has
\beq
[\boldsymbol{\widetilde{\gamma}}^{-1}]_{IJ}\cdot
~\eps^{-|X_{\overline{N^I}}|-|X_{\overline{N^J}}|}~=~
[\boldsymbol{\widetilde{\gamma}}^{-1}]_{IJ}\cdot
~\eps^{X_{\overline{N^I}}+X_{\overline{N^J}}},
\eeq
the mass matrix of the light neutrinos reads
\beq
\frac{\langle H^{\mcal{U}} \rangle^2}{m_{3/2}}~~\eps^{X_{L^i}+2X_{H^{\mcal{U}}}+X_{L^l}}~\sum_{J,K} \eps^{2(X_{\overline{N^J}}+X_{\overline{N^K}})}~\approx~\frac{\langle H^\mcal{U} \rangle^2}{m_{3/2}}~~\eps^{X_{L^i}+2X_{H^\mcal{U}}+X_{L^l}-2n},
\eeq
with
\beq
n~\equiv~2~\mbox{max}\big\{|X_{\overline{N^1}}|,
|X_{\overline{N^2}}|\big\}.
\eeq
\emph{So, unlike the corresponding expressions for  positive  
$X_{\overline{N^I}}$ in cases \emph{1.)} and \emph{2.)},  
here  the $X_{\overline{N^I}}$ do not drop out.} 
Analogous to the reasoning in case \emph{2.)} we get 
(with the lightest neutrino again  without  mass)
\beq
\mbox{$\frac{19.106}{2}$}\approx X_{L^2}+X_{H^\mcal{U}}-n~~~\mbox{and}~~~
\mbox{$\frac{17.864}{2}$}\approx X_{L^3}+X_{H^\mcal{U}}-n,
\eeq
to be rounded such that $X_{L^i}~+~X_{H^\mcal{U}}=$ half-odd-integer.  So with $X_{L^2}=X_{L^3}$ we get\footnote{If we allow a larger gravitino mass $m_{3/2}
  \simeq 10$ or  $100$~TeV, other possibilities arise, such as 
$17/2+n$
 or
  $15/2+n$, which lead to $\Delta^{\!H} = 10+n$ or $9+n$, 
respectively.  However,
  they tend to give a $\mu$-term which is too large and we will not consider them
  further in this paper. }
\beq\label{diezweite}
X_{L^{2,3}}+X_{H^\mcal{U}}-n=\frac{19}{2}.
\eeq
Thus from Eq.~(\ref{blc}) one gets $
0<-3X_{\overline{N^J}}<\frac{19}{2}+n.$
Thus as long as
\beq
 X_{\overline{N^1}},~ X_{\overline{N^2}}~~\in~~\Bigg\{~-~\frac{1}{2},~-~\frac{3}{2},~-~...,~-~\frac{17}{2}\Bigg\}.
\eeq
any pair of $\{ X_{\overline{N^1}}, X_{\overline{N^2}} \}$ is fine.

Eq.(\ref{diezweite}) leads to 
\beq
\Delta^{\!H}=11+n.
\eeq
The results are displayed in Tables~\ref{Table7}, \ref{Table8}. 
Only the combinations $\{x=3,n=1\}$, $\{x=2,n=1\}$, $\{x=2,n=3\}$ 
give an
$X$-charge assignment with a   maximum absolute value smaller than 10.   $\{x=3,n=1\}$ is particularly nice because the denominators of the
$X$-charges are given by $5,10$ or $20$.     
\begin{table}
\begin{center}
\begin{tabular}{|c|}
\hline $\phantom{\Bigg|}-\frac{19}{2}< 
X_{\overline{N^1}}~<~0,~~~~$  $\phantom{\Bigg|}-
\frac{19}{2}< X_{\overline{N^2}}~<~0,~~~~$  \\
\hline
\end{tabular}~
\end{center}
\begin{center}
\begin{tabular}{|c|}
\hline $\phantom{\Bigg|}X_{H^\mathcal{D}}~=~
-~\frac{357+38n}{90},~~~~$  $X_{H^\mathcal{U}}=
~\frac{267+38n}{90}$  \\
\hline
\end{tabular}~
\end{center}
\begin{center}
\begin{tabular}{|c|c|c|c|c|c|}
\hline
\bf{Generation}\phantom{$\Big|$} $\bsym{i}$~ &~~~ $
\bsym{X_{\!Q^i}}$~~~ &~~~ $\bsym{X_{\!\overline{D^i}}}$
~~~ &~~~ $
\bsym{X_{\!\overline{U^i}}}$~~~ &~~~ $\bsym{X_{\!L^i}}$
~~~ &~~~ $
\bsym{X_{\!\overline{E^i}}}$~~~       \\\hline
1\phantom{\Bigg|} & $\frac{291-26n}{135}$   &$\frac{703}{90}+\frac{83n}{135}$ 
&$\frac{777-62n}{270}$   &$\frac{339+26n}{45} $ &$\frac{309-14n}{90} 
$  \\  \hline
2\phantom{\Bigg|} & $\frac{156-26n}{135}$  &$\frac{613}{90}+\frac{83n}{135}$ 
&$-\frac{33+62n}{270}\phantom{-}$&$\frac{294+26n}{45} $ &
$\frac{129-14n}{90} $  \\  \hline 
3\phantom{\Bigg|} & $-\frac{114+26n}{135}\phantom{-}$ &
$\frac{613}{90}+\frac{83n}{135}$ &$-\frac{573+62n}{270}\phantom{-}$ &
$\frac{294+26n}{45} $ &$-\frac{51+14n}{90}\phantom{-}  
$ \\  \hline 
\end{tabular}~.
\end{center}
\caption{Solution to case \emph{6.)} ($\Delta^{\!H}=11+n$) 
with $x=2$.}
\label{Table7}
\end{table}
\begin{table}
\begin{center}
\begin{tabular}{|c|}
\hline $\phantom{\Bigg|}-\frac{19}{2}< 
X_{\overline{N^1}}~<~0,~~~~$  $\phantom{\Bigg|}-
\frac{19}{2}<X_{\overline{N^2}}~<~0,~~~~$  \\
\hline
\end{tabular}~
\end{center}
\begin{center}
\begin{tabular}{|c|}
\hline $\phantom{\Bigg|}X_{H^\mathcal{D}}~=~
-~\frac{353+42n}{100},~~~~$  $X_{H^\mathcal{U}}=
~\frac{253+42n}{100}$  \\
\hline
\end{tabular}~
\end{center}

\begin{center}
\begin{tabular}{|c|c|c|c|c|c|}
\hline
\bf{Generation}\phantom{$\Big|$} $\bsym{i}$~ &~~~ $
\bsym{X_{\!Q^i}}$~~~ &~~~ $\bsym{X_{\!\overline{D^i}}}$
~~~ &~~~ $
\bsym{X_{\!\overline{U^i}}}$~~~ &~~~ $\bsym{X_{\!L^i}}$
~~~ &~~~ $
\bsym{X_{\!\overline{E^i}}}$~~~       \\\hline
1\phantom{\Bigg|} & $\frac{703-58n}{300}$             
&$\frac{614+46n}{75}$ &$\frac{469-34n}{150}$            
&$\frac{797+58n}{100} $ &$\frac{89-4n}{25} $  \\  \hline
2\phantom{\Bigg|} & $\frac{403-58n}{300}$             
&$\frac{539+46n}{75}$ &$\frac{19-34n}{150}$             
&$\frac{697+58n}{100} $ &$\frac{ 39-4n}{25} $  \\  \hline 
3\phantom{\Bigg|} & $-\frac{197+58n}{300}\phantom{-}$ 
&$\frac{539+46n}{75}$ &$-\frac{281+34n}{150}\phantom{-}$ 
&$\frac{697+58n}{100} $ &$-\frac{11+4n}{25}\phantom{-}  
$ \\  \hline 
\end{tabular}~.
\end{center}
\caption{Solution to case \emph{6.)} ($\Delta^{\!H}=11+n$) with 
$x=3$.}
\label{Table8}
\end{table}
%

{}From the Tables~\ref{Table7}, \ref{Table8}  we get 
furthermore that $3\Delta^{\!H}+\sum_IX_{\overline{N^I}}$
 does not allow for Green-Schwarz anomaly cancellation of
$\mathcal{A}_{GGX}$ [\emph{c.f.}   Eq.~(\ref{gvsc}) and the text below it]. So again we are
 forced to require the existence of at least one 
$X$-charged superfield in the hidden   sector.\\

Pseudo Dirac neutrinos are  possible as in  \emph{2.)},
if one has large $X_{\overline{N^I}}$.  But since the 
mass matrix
of the right-handed neutrinos is  $\mathcal{O}(m_{3/2})$ 
rather than $\mathcal{O}(M_{grav})$ as in \emph{2.)},
the $X_{\overline{N^I}}$ do not have to be as large as
in   \emph{2.)},  but one still needs 
$X_{\overline{N^{I}}}<-10$ so that $X_{L^i}+X_{H^\mathcal{U}}>31$,
[\emph{c.f.}  Eq.~(\ref{twenty-one})].
  We shall not pursue this 
idea further.\\

%
%
%
%
\section{\label{soc}Summary, Conclusion and Outlook}
\cleqn

We have constructed a viable theory of flavor
based on a minimal set of ingredients: the anomalous $U(1)_X$
inspired by string theory, only two mass scales, $M_{grav}$
and $m_{3/2}$, one flavon, and two right-handed neutrinos.  It
explains the masses and mixings of quarks, leptons, and neutrinos, the
origin of conserved $R$-parity, and the longevity of the proton.  Note that the
mass scale of the right-handed neutrinos is determined also from
$M_{grav}$ and the $U(1)_X$ symmetry, unlike most models in the literature
that assume a separate origin of their mass scale.

We presented four viable sets of $X$-charges in Tables~\ref{Table5},
\ref{Table6}, \ref{Table7}, and \ref{Table8}. Many of these
$X$-charges are esthetically not pleasing, \emph{i.e.}  highly fractional.
But it should be pointed out that the few models which were found to
be compatible with the bounds on exotic processes in Ref.~\cite{dt}
(without imposing $R$-parity by hand) all needed large or very
fractional $X$-charges, too.  Furthermore, superstring phenomenology
by no means predicts that at low energies one should have moderate or
even easy fractions. As an example, see the (non-anomalous)
beyond-$S\!M$ charges in Ref.~\cite{9807479}.


In particular, the $U(1)_X$ charge assignment in Table~\ref{Table5}
appears esthetically most pleasing, and its choice $x=2$ makes the
resulting proton decay rate an excellent target for future
experiments.

It is quite likely that the $X$-charge assignments can be drastically
improved.  Even though the anomaly cancellation conditions have to
hold exactly, the phenomenological ans\"atze for mass matrices
Eqs.~(\ref{fihr}--\ref{droi}) are surely approximate.  Furthermore, we
did not pursue other phenomenologically viable patterns of mass
matrices, Eqs.~(\ref{fihr},\ref{funef},\ref{droi}).  It would be very
interesting to see if other patterns would lead to much more
attractive $U(1)_X$ charge assignments.

It would be also interesting to check the validity of the models
presented in Tables~\ref{Table5}-\ref{Table8} by a statistical
treatment of the type demonstrated in Figure~1.  Furthermore, we have
not investigated the issue of leptogenesis  in this paper.

Last but not least, it is tempting to repeat the calculations
of this paper for $B_3$, see Eq.~(\ref{be-drei}), and so-called proton 
hexality ($P_6$), see Ref.~\cite{Dreiner:2005rd}, instead of $R_p$:
Both would render the proton stable (forbidding and not just suppressing 
$QQQL$), while the former allows the necessary $R_p$-violating operators 
to generate neutrino masses without having to introduce 
right-handed neutrinos. We come back to  these points 
in separate papers, see Refs.~\cite{Dreiner:2006xw,dlmtII}.

%
%
%
%
\section{Acknowledgments}
%
We thank Kaladi S. Babu, Mirjam Cveti{\v{c}}, Alon E. Faraggi,
Nobuhiro Maekawa and Steve Martin for useful correspondence and 
Lisa Everett,  Howie Haber, Alejandro Ibarra, Tim Jones, Christoph Luhn  
and Hans-Peter Nilles 
for helpful conversation. 
Daniel Larson pointed out 
several typos.
 M.T. greatly appreciates that his work was
 supported by a
fellowship within the Postdoc-Programme of the German Academic
Exchange Service (Deutscher Akademischer Austauschdienst, DAAD).
  The work of H.M. was supported by the Institute for Advanced Study, funds for
Natural Sciences, as well as in part by the DOE under contracts
DE-AC03-76SF00098 and in part by  NSF grant PHY-0098840.

%
%
%
%

\begin{appendix}

%
%
%
%
\section{\label{seeesooo}The Seesaw  Mechanism}
\cleqn

We shall slightly generalize a calculation from 
Ref.~\cite{branco} for the special case $X_{L^i}+
X_{H^\mathcal{U}},~X_{\overline{N^I}}>0$. The leptonic 
sector contains the fermionic mass terms given
 in Eq.~(\ref{notbigbookbutbigequation}), \emph{i.e.} after 
$H^\mathcal{U}$ has acquired a VEV we have 
\beqn
\frac{1}{2}~~\left(\bsym{n_L}^T, ~~\overline{\bsym{n_R}}^T
 \right)\cdot\left(\begin{array}{ccc}\frac{\langle 
H^{\mathcal{U}}\rangle^2}{M_{grav}}~\bsym{\Psi} &~&
 {\langle H^{\mathcal{U}}\rangle}~\bsym{G^{(N)}}\\ 
~&~&~\\ \langle H^{\mathcal{U}}\rangle~{\bsym{G^{(N)}}^T}
 &~&M_{grav}~\bsym{\Gamma}\end{array}\right)\cdot
\left(\begin{array}{c}\bsym{n_L} \\ ~\\ \overline{
\bsym{n_R}} \end{array} \right).
\eeqn
Now as shown below we insert two unit matrices. They  are each  products of
two  unitary matrices, in such a way  that  the above  mass
 matrix  is Schur diagonalized, 
with all resulting  entries being non-negative. 
The diagonal entries 
are the neutrino masses. They are equal to the 
 positive square roots of the
eigenvalues of the above mass matrix times its adjoint. Using 
%
\beq
\eta\equiv\frac{\langle H^{\mathcal{U}}\rangle}{M_{grav}}\ \ ,
\eeq
we find
\beq\label{h-zwei}
\frac{M_{grav}}{2}~\left(\bsym{n_L}^T, ~~\overline{
\bsym{n_R}}^T \right)\cdot\underbrace{\bsym{U_{n}}^T~
\bsym{U_{n}}^{*}}_{\bsym{1\!\!1}}\cdot\left(
\begin{array}{ccc} \eta^2~\bsym{\Psi} &~& {\eta}~{{
\bsym{G^{(N)}}}}\\ ~&~&~\\{\eta}~{\bsym{G^{(N)}}^T} &~&
\bsym{\Gamma}\end{array}\right)\cdot\underbrace{\bsym{
U_{n}}^\dagger~\bsym{U_{n}}}_{\bsym{1\!\!1}}\cdot
\left(\begin{array}{c}\bsym{n_L} \\ ~\\ \overline{
\bsym{n_R}} \end{array} \right).~~~~~
\end{equation}
Let us write
\beq\label{sowasvonkeinenbockmehr}
\bsym{U_n}\equiv \left(\begin{array}{ccc} \bsym{V}_{11}
 &~& \bsym{V}_{12}\\ ~&~&~\\ \bsym{V}_{21} &~&  
\bsym{V}_{22} \end{array}\right),
\eeq
So, $\bsym{D^{(...)}}$ being diagonal ($\bsym{\nu},\bsym{\omega}$ denote mass eigenstates),
\beqn\label{se}
\left(\begin{array}{ccc} \bsym{D^{(\nu)}}  & ~&  \\ 
~&~&~\\  &~&  \bsym{D^{(\omega)}}\end{array}\right)=
~~~~~~~~~~~~~~~~~~~~~~~~~~~~~~~~~~~~~~~~~~~~~~~~~~~~~~
~~~~~~~~~~~~~~\nonumber\\~~~~~~~~~~~~~~~~~~~~~~~~~~~~~~
~~~~~~~~~~~~~~~~~~~~~~~~~~~~~~~~~~~~~~\nonumber\\
\left(\begin{array}{ccc} {\bsym{V}_{11}}^* &~& {
\bsym{V}_{12}}^*\\ ~&~&~\\ {\bsym{V}_{21}}^* &~&  {
\bsym{V}_{22}}^* \end{array}\right)\cdot\left(
\begin{array}{ccc} \eta^2~\bsym{\Psi}  &~& \eta~ {
\bsym{G^{(N)}}} \\ ~&~&~\\ \eta~ {{\bsym{G^{(N)}}}^T} 
&~&  \bsym{\Gamma} \end{array}\right)\cdot\left(
\begin{array}{ccc} {\bsym{V}_{11}}^\dagger &~& {
\bsym{V}_{21}}^\dagger\\ ~&~&~\\ {\bsym{V}_{12}}^\dagger
 &~&  {\bsym{V}_{22}}^\dagger \end{array}\right),
\eeqn
so that
\beqn\label{seaso1}
{\bsym{V}_{12}}^*~\bsym{\Gamma}~{\bsym{V}_{12}}^\dagger~+~ 
    \eta~  {\bsym{V}_{12}}^*~{{\bsym{G^{(N)}}^T}}~{
\bsym{V}_{11}}^\dagger~+~ 
    \eta~  {\bsym{V}_{11}}^*~{{\bsym{G^{(N)}}}}~{
\bsym{V}_{12}}^\dagger~ +~\eta^2~{\bsym{V}_{11}}^*~
\bsym{\Psi}~{\bsym{V}_{11}}^\dagger&=&\bsym{D^{(\nu)}},
~~~~~~\nonumber\\  & & \\\label{seaso2}
{\bsym{V}_{22}}^*~   \bsym{\Gamma}~{\bsym{V}_{22}}^\dagger
~ +  ~\eta~{\bsym{V}_{22}}^*~ {{\bsym{G^{(N)}}^T}}~{
\bsym{V}_{21}}^\dagger~+~   
   \eta~  {\bsym{V}_{21}}^*~{{\bsym{G^{(N)}}}}~{
\bsym{V}_{22}}^\dagger~ +~\eta^2~{\bsym{V}_{21}}^*~
\bsym{\Psi}~{\bsym{V}_{21}}^\dagger  &=&\bsym{D^{(
\omega)}},~~~~~~\nonumber\\  & &\\\label{seaso3}
  {\bsym{V}_{22}}^*~  \bsym{\Gamma}~   {
\bsym{V}_{12}}^\dagger~  +~ \eta ~ {\bsym{V}_{22}}^*~ 
 {{\bsym{G^{(N)}}^T}}   ~ {\bsym{V}_{11}}^\dagger~ +~
  \eta~  {\bsym{V}_{21}}^*  ~ {{\bsym{G^{(N)}}}}~ 
{\bsym{V}_{12}}^\dagger    ~ +
    ~  \eta^2~  {\bsym{V}_{21}}^*~ \bsym{\Psi}~     
{\bsym{V}_{11}}^\dagger       &=&  0.\nonumber\\  & & 
\eeqn
Now in the limiting case where $\eta=0$ we have instead
 of Eq.~(\ref{se})
\beq
\left(\begin{array}{ccc} 0 & ~&  \\ ~&~&~\\  &~&  
\bsym{D^{(\omega)}}\end{array}\right)=\left(
\begin{array}{ccc} {\bsym{V}_{12}}^*~ \bsym{\Gamma}~ 
 {\bsym{V}_{12}}^\dagger  &~&{\bsym{V}_{12}}^*~ 
\bsym{\Gamma}~  {\bsym{V}_{22}}^\dagger \\ ~&~&~\\ 
{\bsym{V}_{22}}^*~ \bsym{\Gamma}~  {
\bsym{V}_{12}}^\dagger  &~&{\bsym{V}_{22}}^* ~
\bsym{\Gamma} ~ {\bsym{V}_{22}}^\dagger  \end{array}
\right),
\eeq
and in this case we   need ${\bsym{V}_{12}}=0$. Taking this into account we arrive at
\beq
\bsym{U_n}^{-1}=\left(\begin{array}{ccc} {
\bsym{V}_{11}}^{-1} &~& 0\\ ~&~&~\\  -{
\bsym{V}_{22}}^{-1}~{\bsym{V}_{21}}~ {
\bsym{V}_{11}}^{-1} &~& {\bsym{V}_{22}}^{-1} 
 \end{array}\right),
\eeq
which has to  equal 
\beq
\bsym{U_n}^{\dagger}=\left(\begin{array}{ccc} 
{\bsym{V}_{11}}^\dagger &~& {\bsym{V}_{21}}^\dagger\\
 ~&~&~\\  0 &~& {\bsym{V}_{22}}^{\dagger}  \end{array}
\right).
\eeq
Hence we also  need ${\bsym{V}_{21}}=0$. This 
little exercise demonstrates that  
for $\eta\ll1$   the deviation of  
${\bsym{V}_{11}},~{\bsym{V}_{22}}$ from being unitary 
is  $\mathcal{O}(\eta)$.  Furthermore, ${\bsym{V}_{12}},
~{\bsym{V}_{21}}$ are suppressed by a factor of  $\eta$
 compared to  ${\bsym{V}_{11}},~{\bsym{V}_{22}}$ (hence
 $\bsym{n_L}={\bsym{V}_{11}}^\dagger~\bsym{\nu}~+~
{\bsym{V}_{21}}^\dagger~\bsym{\omega}$ can be approximated
 as $\bsym{n_L}\approx{\bsym{V}_{11}}^\dagger~\bsym{\nu}$).
 Writing 
\beq\label{h-drei}
{\bsym{V}_{12}}=\eta~
{{\bsym{V}_{12}}}^\prime,~~~{\bsym{V}_{21}}=
\eta~{{\bsym{V}_{21}}}^\prime
\eeq 
and dropping higher 
orders of $\eta$, Eq.~(\ref{seaso3})  can be approximated as
\beq
{\bsym{V}_{22}}^*~  \bsym{\Gamma}~   {{
\bsym{V}_{12}}^\prime}^\dagger~  +~ {\bsym{V}_{22}}^*~  
{{\bsym{G^{(N)}}^T}}  ~ {\bsym{V}_{11}}^\dagger~=~0,
\eeq
thus
\beq
\bsym{\Gamma}~   {{\bsym{V}_{12}}^\prime}^\dagger~  
+~ {{\bsym{G^{(N)}}^T}}  ~ {\bsym{V}_{11}}^\dagger~=~0.
\eeq
Assuming that $\bsym{\Gamma}$ is non-singular one finds
\beq
{{\bsym{V}_{12}}^\prime}^\dagger~=~-~\bsym{\Gamma}^{-1}~
 {{\bsym{G^{(N)}}^T}}   ~ {\bsym{V}_{11}}^\dagger.
\eeq
Inserting this into  Eq.~(\ref{seaso1}) leads to the masses 
of the light neutrinos: the diagonal elements of 
\beqn
\frac{\langle H^\mathcal{U} \rangle^2}{M_{grav}}~~
{\bsym{V}_{11}}^*~~\Big(~{\bsym{\Psi}~-~{
\bsym{G^{(N)}}}~\bsym{\Gamma}^{-1}~\bsym{G^{(N)}}^T} ~  
\Big) ~~ {\bsym{V}_{11}}^\dagger.
\eeqn
Note that this holds for an arbitrary number of 
$\overline{N}$, as was  used \emph{e.g.} in 
Ref.~\cite{faraggithormeier}.
%

%
%
%
\section{\label{kalk}Relating Masses with Powers of 
${\bsym{\eps}}$}
\cleqn

How to extract quark and charged lepton masses
from hierarchical matrices $\bsym{G^{(U,D,E)}}$ is 
demonstrated in Ref.~\cite{dt}.
\emph{E.g.}  $\bsym{G^{(U)}}$ without (filled up) 
supersymmetric zeros
gives masses proportional to 
${G^{(U)}}_{11}\sim\eps^{X_{Q^1}+
X_{H^\mathcal{U}}+{X_{\overline{U^1}}}}$, 
${G^{(U)}}_{22}\sim\eps^{X_{Q^2}+
X_{H^\mathcal{U}}+{X_{\overline{U^2}}}}$, 
${G^{(U)}}_{33}\sim\eps^{X_{Q^3}+
X_{H^\mathcal{U}}+{X_{\overline{U^3}}}}$.
The situation is slightly different for 
the $\bsym{{\mathcal{G}^{(\nu)}}}$ of  \emph{2.)} in 
Subsection~\ref{921} because of its vanishing determinant. 
We assume  that there is enough
hierarchy in $\bsym{{\mathcal{G}^{(\nu)}}}$  such that 
the absolute values of the eigenvalues $\lambda$  of 
$\bsym{{\mathcal{G}^{(\nu)}}} $ are approximately the
 positive square roots  of  the eigenvalues of   
$\bsym{{\mathcal{G}^{(\nu)}}}\bsym{{
\mathcal{G}^{(\nu)}}}^\dagger$. It follows that it 
suffices to examine the characteristic polynomial of 
$\bsym{{\mathcal{G}^{(\nu)}}}$:
\beqn
\lambda^3~-~\mbox{tr}[\bsym{\mathcal{G}^{(\nu)}} ]~
\lambda^2~+~\frac{1}{2}~\Big(\mbox{tr}[\bsym{
\mathcal{G}^{(\nu)}} ]^2~-~\mbox{tr}[\bsym{
\mathcal{G}^{(\nu)}}\bsym{\mathcal{G}^{(\nu)}} ]  \Big)~
\lambda~-~\underbrace{\det[\bsym{\mathcal{G}^{(\nu)}}
]}_{=~0},
\eeqn
thus we are interested in 
\beqn
&&\lambda^2~-~\mbox{tr}[\bsym{\mathcal{G}^{(\nu)}} ]~
\lambda~+~\frac{1}{2}~\Big(\mbox{tr}[\bsym{
\mathcal{G}^{(\nu)}} ]^2~-~\mbox{tr}[\bsym{
\mathcal{G}^{(\nu)}}\bsym{\mathcal{G}^{(\nu)}} ]  
\Big)~=~0.~~~~~
\eeqn
Now 
\beq
\mbox{tr}[\bsym{\mathcal{G}^{(\nu)}} ]~\sim~\mbox{max}
\big\{{\mathcal{G}^{(\nu)}}_{11},~{
\mathcal{G}^{(\nu)}}_{22},~{\mathcal{G}^{(\nu)}}_{33}\big\}
,
\eeq
and  
\beqn
\frac{1}{2}~\Big(\mbox{tr}[\bsym{\mathcal{G}^{(\nu)}} 
]^2~-~\mbox{tr}[\bsym{\mathcal{G}^{(\nu)}}\bsym{
\mathcal{G}^{(\nu)}} ] \Big)~\sim~~~~~~~~~~
~~~~~~~~~~~~~~~~~~~~~~~~~~~~~~~~~~~~~~\nonumber\\\mbox{max}
\big\{ {\mathcal{G}^{(\nu)}  }_{11},
~{\mathcal{G}^{(\nu)}}_{22},~{\mathcal{G}^{(\nu)}}_{33}\}
~\times~\mbox{middle}\big\{{\mathcal{G}^{(\nu)}}_{11},~ 
{\mathcal{G}^{(\nu)}}_{22},~{\mathcal{G}^{(\nu)}}_{33}\},
\eeqn
so that 
\beq
\mbox{tr}[\bsym{\mathcal{G}^{(\nu)}} ]>\frac{1}{2}~
\Big(\mbox{tr}[\bsym{\mathcal{G}^{(\nu)}} ]^2~-~
\mbox{tr}[\bsym{\mathcal{G}^{(\nu)}}\bsym{
\mathcal{G}^{(\nu)}} ]  \Big).
\eeq
So the largest eigenvalue   is of the order of 
\beq
\mbox{max}\big\{{\mathcal{G}^{(\nu)}}_{11},~
{\mathcal{G}^{(\nu)}}_{22},~{\mathcal{G}^{(\nu)}}_{33}
\big\},
\eeq
and the other non-zero eigenvalue  is of the order of
\beq
\mbox{middle}\big\{{\mathcal{G}^{(\nu)}}_{11},
~{\mathcal{G}^{(\nu)}}_{22},~{\mathcal{G}^{(\nu)}}_{33}
\big\}.
\eeq

 %
 %
 %
 \section{\label{mitsusy0}Including 
 Supersymmetric Zeros}
\cleqn

 We are now going to investigate  the case with the  
$X$-charges of all right-handed neutrino superfields 
 being positive, however we allow for a few supersymmetric
 zeros in $\bsym{G^{(N)}}$ and  $\bsym{G^{(E)}}$,
generalizing Section~\ref{921} (see however the \emph{caveat}
mentioned in Section \ref{pHeNo}).
 From 
$\langle H^\mathcal{U}\rangle^2/M_{grav}\ll\sqrt{
3\times10^{-3~}}~$eV we get that $X_{L^i}+X_{
H^\mathcal{U}}<0$, which is again  why we   do not get 
any  substantial contribution from  $LH^\mathcal{U} L
 H^\mathcal{U}$. Expressing the mass matrix of the 
light neutrinos  in terms of the coupling constants 
which we have before canonicalizing the K\"ahler potential
gives
\beqn\label{preckkk}
\bsym{\mathcal{G}^{(\nu)}}=
\frac{{\bsym{C^{(L)}}^{-1}}^T}{\sqrt{H^{(H^\mathcal{U})}~}}
~~\underbrace{~{\bsym{G^{(N)}}}_{preCK}~{\bsym{
\Gamma}_{preCK}}^{-1}~{\bsym{G^{(N)}}_{
preCK}}^T ~}_{\equiv \bsym{{\mathcal{G}^{(\nu)}}}
_{preCK}}~~
\frac{\bsym{C^{(L)}}^{-1}}{
\sqrt{H^{(H^\mathcal{U})}~}},~~~~
\eeqn
the $\bsym{C^{(\overline{N})}}$ in Eq.~(\ref{preckkk}) 
having 
mutually canceled each other. With
\beqn
{{G^{(N)}}_{\!preCK~}}_{iJ}&=&\underbrace{{{
g^{(N)}}_{iJ}}~~\Theta[X_{L^i}+X_{H^\mathcal{U}}+X_{
\overline{N^J}}]}_{\equiv ~~\widehat{{g^{(N)}}}_{iJ}}
~~\eps^{X_{L^i}+X_{H^\mathcal{U}}+X_{\overline{N^J}}  },
\\
~\nonumber\\
{\Gamma_{\!preCK~}}_{IJ}&=&{\gamma_{IJ}}~~\eps^{
X_{\overline{N^I}} +X_{\overline{N^J}}  }
\eeqn
we have, introducing  $\bsym{{g^{(\nu)}}}_{preCK}$,
\beqn
{{{\mathcal{G}^{(\nu)}}}_{preCK~}}_{ij}~=~
\epsilon^{X_{L^i}+X_{L^j}+2X_{H^\mathcal{U}}}~~
\underbrace{\sum_{K,L}~~ \widehat{g^{(N)}}_{iK}~~
[\bsym{{\gamma}}^{-1}]_{KL}~~\widehat{g^{(N)}}_{jL}}_{
\equiv {{g^{(\nu)}}_{preCK}}_{~ij}}.
\eeqn
So, 
what kind of  $\bsym{{g^{(\nu)}}}_{preCK}$
 does one get? If  
$\bsym{\widehat{g^{(N)}}}$ has 
\begin{enumerate}
\item zero supersymmetric zeros, then $\bsym{ 
{g^{(\nu)}}}_{preCK}$ has no textures and two eigenvalues 
$\neq0$; this is the case which we examined in 
detail in Section~\ref{921} \emph{2.)},
\item one  supersymmetric zero (six different 
possibilities), then $\bsym{ {g^{(\nu)}}}_{preCK}$ has no 
textures and two eigenvalues $\neq0$,
\item two  supersymmetric zeros in the same column 
(six different possibilities), then $\bsym{ 
{g^{(\nu)}}}_{preCK}$ has no textures and two eigenvalues 
$\neq0$,
\item two  supersymmetric zeros  in the same row 
(three different possibilities), then $\bsym{ {
g^{(\nu)}}}_{preCK}$ has five textures, so to speak 
``second-generation supersymmetric zeros'' 
(such that there is a non-zero $2\times2$ 
submatrix on the diagonal) and two eigenvalues 
$\neq0$,
\item three  supersymmetric zeros 
\emph{not} all in the same column but two of 
them in the same row  (twelve different possibilities),
 then $\bsym{ {g^{(\nu)}}}_{preCK}$ has five textures 
(such that there is a non-zero $2\times2$ submatrix 
on the diagonal) and two eigenvalues $\neq0$,
\item three supersymmetric zeros  all in the same 
column (two different possibilities), then $\bsym{ 
{g^{(\nu)}}}_{preCK}$ has no textures  but only 
one eigenvalue $\neq0$,
\item four supersymmetric zeros with the two non-zero 
entries being in the same column (six different 
possibilities), then $\bsym{ {g^{(\nu)}}}_{preCK}$ has five 
textures (such that there is a non-zero $2\times2$ 
submatrix on the diagonal) and one eigenvalue $\neq0$,
\item four supersymmetric zeros with the two non-zero 
entries being in the same row (three different 
possibilities), then $\bsym{ {g^{(\nu)}}}_{preCK}$ has eight 
textures (such that there is a non-zero entry on the 
diagonal) and one eigenvalue $\neq0$,
\item five supersymmetric zeros (six different 
possibilities), then $\bsym{{g^{(\nu)}}}_{preCK}$
 has eight 
textures (such that there is one non-zero entry  on the 
diagonal) and one eigenvalue $\neq0$,
\item six supersymmetric zeros, then $\bsym{ {
g^{(\nu)}}}_{preCK}$ has nine textures  
and no eigenvalue $\neq0$.
\end{enumerate}

Now
\begin{itemize} 
\item Clearly the  $X$-charges have to be such that 
Points~6., 7., 8., 9., and 10. are forbidden. 
\item With $\frac{{{C^{(L)}}^{-1}}_{ij}}{\sqrt{H^{(
H^\mathcal{U})}}}\sim\eps^{|X_{L^i}-X_{L^j}|}$,
the order of magnitude of suppression of the individual 
entries of Points~1., 2., and 3. is not affected by 
the canonicalization of the K\"ahler potential, so
that 
$\bsym{\mathcal{G}^{(\nu)}}\sim\bsym{
\mathcal{G}^{(\nu)}}_{{preCK}}$:  In 
Ref.~\cite{blr} it was shown that with no supersymmetric 
zeros we have \emph{e.g.} that $\bsym{G^{(U)}}$ remains 
unchanged 
(concerning the $\epsilon$-suppression of the individual
 entries):
\beqn\sum_{j,k}\epsilon^{|X_{Q^i}-X_{Q^j}|}\cdot
\epsilon^{X_{Q^j}+X_{H^{\mathcal{U}}}+X_{\overline{U^k}}}
\cdot\epsilon^{|X_{\overline{U^k}}-X_{\overline{U^l}}|}~
\approx~\epsilon^{X_{Q^i}+X_{H^{\mathcal{U}}}+
X_{\overline{U^l}}}.
\eeqn
Now  make the replacements $X_{Q^i}\rightarrow 
X_{L^i},~X_{\overline{U^i}}\rightarrow X_{L^i}+
X_{H^{\mathcal{U}}}$, and thus
\beqn
\sum_{j,k}\epsilon^{|X_{L^i}-X_{L^j}|}  \cdot
\epsilon^{X_{L^j}+2~X_{H^{\mathcal{U}}}+X_{{{L^k}}}}
\cdot\epsilon^{|X_{L^i}-X_{L^j}|}\approx~
\epsilon^{X_{L^i}+2X_{H^{\mathcal{U}}}+X_{{L^l}}}.
\eeqn
This in hindsight justifies that in Section~\ref{921}
 we could afford not to explicitly perform a proper 
canonicalization of the K\"ahler potential. So we can 
slightly relax the result presented in 
Tables~\ref{Table5} and \ref{Table6}: Instead 
of $X_{\overline{N^{1,2}}}\leq5/2$  we may have  
$X_{\overline{N^1}}\leq 3/2$, $X_{\overline{N^1}}\leq 5/2$
 or $X_{\overline{N^2}}\leq 5/2$, $X_{\overline{N^1}}
\leq3/2$ (Points~1. and 3.).
\item For the Points~4. and 5. the effects of 
the  canonicalization  
are  more elaborate.  Take \emph{e.g.}  ($\bsym{\times}$ 
symbolizes  any  non-zero entry) 
\beq
\bsym{G^{(N)}}_{preCK}=\left(\begin{array}{cc}
 \!\!0 & \!\!0\\\!\!\bsym{\times} & \!\!\bsym{\times} 
\\\!\!\bsym{\times} & \!\!\bsym{\times} 
\end{array}\!\!\right);~~\left(\begin{array}{cc} \!\!0 
& \!\!0\\\!\!0 & \!\!\bsym{\times} \\\!\!\bsym{\times} 
& \!\!\bsym{\times} \end{array}\!\!\right),\left(
\begin{array}{cc} \!\!0 & \!\!0\\\!\!\bsym{\times} 
& \!\!0 \\\!\!\bsym{\times} & \!\!\bsym{\times}
 \end{array}\!\!\right),\left(\begin{array}{cc} \!\!0 
& \!\!0\\\!\!\bsym{\times} & \!\!\bsym{\times} \\\!\!0 &
 \!\!\bsym{\times} \end{array}\!\!\right), \left(
\begin{array}{cc} \!\!0 & \!\!0\\\!\!\bsym{\times} 
& \!\!\bsym{\times} \\\!\!\bsym{\times} & \!\!0 
\end{array}\!\!\right).
\eeq 
These come from $X_{L^1}+X_{H^\mathcal{U}}+
X_{\overline{N^1}},~X_{L^1}+X_{H^\mathcal{U}}+
X_{\overline{N^2}}<0$, but all or all but one  of 
 $X_{L^{i\neq1}}+X_{H^\mathcal{U}}+X_{\overline{N^j}}$
 have to be $\geq0$. Dropping higher orders
 of $\eps$ and ignoring $\mathcal{O}(1)$ prefactors, we find
\beqn\bsym{\mathcal{G}^{(\nu)}}_{{preCK}}&\sim&
{\eps^{2X_{H^\mathcal{U}}}~\left(\begin{array}{ccc}  0
  & 0 & 0  \\0  & \eps^{X_{L^2}+X_{L^2}}  & 
\eps^{X_{L^2}+X_{L^3}}   \\0  &  \eps^{X_{L^3}+X_{L^2}}
  &  \eps^{X_{L^3}+X_{L^3}}\end{array}\right)}.
\eeqn
The canonicalization of the K\"ahler potential yields 
then (again to lowest order, but the determinant still
vanishes) 
\beqn
\bsym{\mathcal{G}^{(\nu)}}~\sim~\eps^{
2X_{H^\mathcal{U}}}\left(\begin{array}{ccc}  
\eps^{\widetilde{X_{L^1}}+\widetilde{X_{L^1}}} &
\eps^{\widetilde{X_{L^1}}+X_{L^2}}  &  
\eps^{\widetilde{X_{L^1}}+X_{L^3}} \\
\eps^{X_{L^2}+\widetilde{X_{L^1}}} & 
\eps^{X_{L^2}+X_{L^2}}  & \eps^{X_{L^2}+X_{L^2}}   \\
\eps^{X_{L^3}+\widetilde{X_{L^1}}} &  \eps^{X_{L^3}
+X_{L^2}}  &  \eps^{X_{L^3}+X_{L^3}}     
\end{array}\right),
\eeqn
with $~~~\widetilde{X_{L^1}}=2X_{L^2}-X_{L^1}~~~$ if 
  $~~~X_{L^2}<X_{L^3}~~~$ and $~~~\widetilde{X_{L^1}}
=2X_{L^3}-X_{L^1}$ ~~~if~~~   $X_{L^3}<X_{L^2}$. 
Analogous results of course hold for a 
$\bsym{\mathcal{G}^{(\nu)}}_{{preCK}}$ of the 
form
\beqn& & \eps^{2X_{H^\mathcal{U}}}~\left(
\begin{array}{ccc}  \eps^{X_{L^1}+X_{L^1}} & 0 & 
\eps^{X_{L^1}+X_{L^3}} \\0  & 0  & 0 \\\eps^{X_{L^3}
+X_{L^1}}  &  0  &   \eps^{X_{L^3}+X_{L^3}} \end{array}
\right),
~~~\nonumber\\& & \eps^{2X_{H^\mathcal{U}}}~\left(
\begin{array}{ccc}  \eps^{X_{L^1}+X_{L^1}} &  
\eps^{X_{L^1}+X_{L^2}} & 0 \\\eps^{X_{L^2}+X_{L^1}} & 
 \eps^{X_{L^2}+X_{L^2}}  & 0  \\0      &  0     & 0 
\end{array}\right).~~~~~
\eeqn
\end{itemize}
We are now ready to discuss the neutrino mass spectrum
 without the assumption of having no supersymmetric 
zeros. The neutrino masses give (as in 
Section~\ref{921},~\emph{2.)}) several possibilities:
\begin{itemize}
\item $X_{L^{2,3}}+X_{H^\mathcal{U}}=-5/2$, thus 
\beq
\zeta=\frac{2\Delta^L_{31}-1}{3},~~~\Delta^{\!H}=
-2-\Delta^L_{31},
\eeq
since $\zeta$ is an integer, we get $\Delta^L_{31}=...,
-4,-1,2,5,...$. The case  $\Delta^L_{31}=-1$ was treated 
in  detail in Section~\ref{921},~\emph{2.)}, belonging to
the categories~1.~and~3. All other 
possibilities are not viable (for the calculation of 
${\boldsymbol{U^{\!M\!N\!S}}}$ we made use of the 
expressions
in Ref.~\cite{hr}, adapted to leptons), the case which
resembles Eq.~(\ref{mmnnss}) most is  
$\Delta^L_{31}=-4$, namely 
\beqn
{\boldsymbol{U^{\!M\!N\!S}}}~\sim~\left(
\begin{array}{lll}
~1 & \eps^2 & \eps^4 \\
\eps^2 & 1 & 1 \\
\eps^2 & 1 & 1 \end{array}\right);
\end{eqnarray}
as an example for the rest consider $\Delta^L_{31}=8$,
 leading to
\beqn
{\boldsymbol{U^{\!M\!N\!S}}}~\sim~\left(\begin{array}{lll}
~1 & \eps^8 & \eps^8 \\
\eps^8 & 1 & 1 \\
\eps^8 & 1 & 1 \end{array}\right).
\end{eqnarray}
\item $X_{L^{3,1}}+X_{H^\mathcal{U}}=-5/2$, thus 
\beq
\Delta^L_{31}=0,~~~\Delta^{\!H}=-2.
\eeq
No value for $\zeta$ yields a sensible 
${\boldsymbol{U^{\!M\!N\!S}}}$.
\item  $X_{L^{1,2}}+X_{H^\mathcal{U}}=-5/2$, thus 
\beq
\zeta=\frac{\Delta^L_{31}-1}{3},~~~\Delta^{\!H}=-2;
\eeq
since $\zeta$ is an integer, we get 
$\Delta^L_{31}=...,4,1,-2,-5,...$, 
none of which gives a  
${\boldsymbol{U^{\!M\!N\!S}}}$ in agreement with 
experiment.
\end{itemize}
%

%
%
\section{\label{bplp}Conserved $\boldsymbol{B_p}$ and  
$\boldsymbol{L_p}$; Guaranteeing all Gauge Invariant Terms}
%
%
For completeness' sake it should be mentioned that 
the same reasoning  to conserve 
$R_p$ as presented in Section~\ref{Consr} can be applied 
to $L_p$ and $B_p$ 
instead.  However, $L_p$ and $B_p$ are not 
free of  discrete gauge anomalies and thus not viable, see 
Ref.~\cite{ibaro,ibanezross}, and   we cannot  
conserve any two of these three parities simultaneously 
by virtue of the $X$-charges. 

Instead of 
Eq.~(\ref{l0.5}) we have  that  
\beqn
X_{\overline{N^1}},X_{L^1}-X_{H^\mathcal{D}}~=~\left\{
\begin{array}{l}
\mbox{half-odd-integer}~~(L_p)\\
\\
\mbox{integer}~~(B_p)\end{array}\right..
\eeqn
Furthermore we get instead of Eq.~(\ref{rrpp})
\beqn
n_L~-~n_{\overline{N}}~-~n_{\overline{E}}&=&2\mathcal{L}~+
~\lambda~~~(L_p),\nonumber\\ 
n_Q~-~n_{\overline{U}}~-~
n_{\overline{D}}&=&2\mathcal{B}~+
~\beta ~~~(B_p).
\eeqn
So 
\beq
X_{total}-\mbox{integer}~=~\left\{
\begin{array}{l} 
(3 X_{Q^1} + X_{L^1}-\mbox{$\frac{3}{2}$})\mathcal{C}
-\frac{\lambda}{2},~~~~~(L_p)\\ \\ (3 X_{Q^1} + X_{L^1})
\mathcal{C},~~~~~(B_p)\end{array}\right. .
\eeq
Considering  $B_p$,  we have that
\beqn
\mathcal{C}=\mbox{even}~\Leftrightarrow~B_p,~~~~~
\mathcal{C}=\mbox{odd}~\Leftrightarrow~\not\!\!B_p.
\eeqn 
So for both $L_p$ and $B_p$  we  find the condition 
\beqn
3 X_{Q^1} + X_{L^1}~=~ \mbox{half-odd-integer}.
\eeqn
Unlike Eq.~(\ref{CASE}) this cannot 
be combined with anomaly cancellation  via the 
Green-Schwarz mechanism, see the end of Section~\ref{anom}: 
A third of an integer cannot 
be half-odd-integer.

Opposed to guaranteeing certain parities due to the 
$X$-charges, we might ask for the conditions such that  
\emph{all} gauge invariant  terms  have an integer $X$-charge.
So instead of 
Eq.~(\ref{l0.5}) we have  
\beqn
X_{\overline{N^1}},X_{L^1}-X_{H^\mathcal{D}}~=~\mbox{integer}.
\eeqn
So 
\beq
X_{total}-\mbox{integer}~=~ 
(3 X_{Q^1} + X_{L^1})~\mathcal{C},
\eeq
giving the condition
\beqn
3 X_{Q^1} + X_{L^1}~=~ \mbox{integer}.
\eeqn
The results of this Section  and a comparison to 
Section~\ref{Consr} are summarized in Table~\ref{nochnentable}.
\begin{table}
\begin{center}\begin{tabular}{|c|c|c|}
\hline
 & \begin{tabular}{c}$\phantom{\Big|}3 X_{Q^1} + X_{L^1}$\\
$=~ \mbox{integer}$
\end{tabular}  
&  \begin{tabular}{c}$\phantom{\Big|}3 X_{Q^1} + X_{L^1}$\\
$=~ \mbox{half-odd-integer}$
\end{tabular} \\
\hline
\begin{tabular}{c}$\phantom{\Big|} X_{L^1}-X_{H^\mathcal{D}},X_{\overline{N^1}}$ \\ $\phantom{\Big|}=~ \mbox{integer}$ \end{tabular} & \begin{tabular}{l}  all gauge invariant  \\ terms have integer \\$X$-charge  \end{tabular} &    \begin{tabular}{l} \\ all gauge invariant\\ $B_p$-even terms  have \\ integer $X$-charge, all \\other  
terms are \\forbidden \\ ~\end{tabular} \\
\hline
 \begin{tabular}{c}$\phantom{\Big|} X_{L^1}-X_{H^\mathcal{D}},X_{\overline{N^1}}$ \\ $\phantom{\Big|}=~ \mbox{half-odd-integer}$ \end{tabular}  &  \begin{tabular}{l} \\ all gauge invariant\\ $R_p$-even terms  have \\ integer $X$-charge, all \\other  
terms are \\forbidden \\~ \end{tabular}   & \begin{tabular}{l} \\ all gauge invariant\\ $L_p$-even terms  have \\ integer $X$-charge, all \\other  
terms are \\forbidden \\~ \end{tabular}  \\
\hline
\end{tabular}\end{center}\caption{Conditions on the 
$X$-charges leading to certain shapes of the superpotential.}
\label{nochnentable}
\end{table}

\end{appendix}

%
%
%

\end{document}